\documentclass[aps,pre,preprint]{revtex4-1}
\usepackage{graphicx}
\usepackage{epstopdf, epsfig}
\usepackage{amsmath}
\usepackage{float}

\begin{document}

%\preprint{}

\title{Propagation of gaseous detonation waves in a spatially inhomogeneous reactive medium}

\author{XiaoCheng Mi}
\affiliation{Department of Mechanical Engineering, McGill University, Montreal, Quebec, Canada}

\author{Andrew J. Higgins}
\email[Corresponding author: ]{andrew.higgins@mcgill.ca}
\affiliation{Department of Mechanical Engineering, McGill University, Montreal, Quebec, Canada}

\author{Hoi Dick Ng}
\affiliation{Department of Mechanical and Industrial Engineering, Concordia University, Montreal, Quebec, Canada}

\author{Charles B. Kiyanda}
\affiliation{Department of Mechanical and Industrial Engineering, Concordia University, Montreal, Quebec, Canada}

\author{Nikolaos Nikiforakis}
\affiliation{Cavendish Laboratory, Department of Physics, University of Cambridge, Cambridge, United Kingdom}



%\thanks{}


\date{\today}

\begin{abstract}
Detonation propagation in a compressible medium wherein the energy release has been made spatially inhomogeneous is examined via numerical simulation.  The inhomogeneity is introduced via step functions in the reaction progress variable, with the local value of energy release correspondingly increased so as to maintain the same average energy density in the medium, and thus a constant Chapman Jouguet (CJ) detonation velocity. A one-step Arrhenius rate governs the rate of energy release in the reactive zones. The resulting dynamics of a detonation propagating in such systems with one-dimensional layers and two-dimensional squares are simulated using a Godunov-type finite-volume scheme.  The resulting wave dynamics are analyzed by computing the average wave velocity and one-dimensional averaged wave structure.  In the case of sufficiently inhomogeneous media wherein the spacing between reactive zones is greater than the inherent reaction zone length, average wave speeds significantly greater than the corresponding CJ speed of the homogenized medium are obtained.  If the shock transit time between reactive zones is less than the reaction time scale, then the classical CJ detonation velocity is recovered.  The spatio-temporal averaged structure of the waves in these systems is analyzed via a Favre averaging technique, with terms associated with the thermal and mechanical fluctuations being explicitly computed.  The analysis of the averaged wave structure identifies the super-CJ detonations as weak detonations owing to the existence of mechanical non-equilibrium at the effective sonic point embedded within the wave structure.  The correspondence of the super-CJ behavior identified in this study with real detonation phenomena that may be observed in experiments is discussed.
\end{abstract}

\maketitle

\section{Introduction}
\label{Sec1}
A substantial collection of experimental evidence revealing that all detonation waves in gaseous mixtures possess a transient, multi-dimensional structure has been uncovered over the past 60 years.\cite{Fickett1979,Lee2008}  The structure consists of triple-point interactions between the leading shock and transverse shock waves that result in a cellular wave front. As a result of this spatially and temporally varying shock front compressing the reactive mixture at different strengths, the distribution of post-shock temperature varies greatly over the detonation cell cycle.  Since the exothermic reactions in gaseous combustible mixtures governed by Arrhenius kinetics are very temperature sensitive, the reaction rates in different regions behind the leading shock may differ by several orders of magnitude.  Although zones of prompt reaction triggered by adiabatic compression may exist in regions of the front consisting of strong Mach stems, weakly-shocked pockets of reactant may not be able to undergo significant exothermic reaction due to their thermal history on the time scale of a detonation.\cite{LeeRadulescu2005, Radulescu2007JFM, Kiyanda2013} Particularly in hydrocarbon fuel mixtures, pockets of compressed, unreacted mixture are observed to be separated from the shock front by the shear layer emanating from the triple points propagating transversely across the front.  These pockets eventually burn out during the cell cycle, likely due to a turbulent flame-like mechanism, releasing their energy in compression waves that still help to support the leading front.  Successive generations of computational simulation of this phenomenon since the late 1970s have revealed greater and greater intricacy of the details, many of which have also been observed in experiments, making a notoriously challenging problem for theoretical description.\cite{Taki1978,Oran1982,Radulescu2007JFM, Mahmoudi2011, Mazaheri2012}\\

While detonation waves in pre-mixed, homogenous media exhibit localized spatio-temporal reaction zone structures (e.g., unburned pockets, etc.), greater complications might arise from spatially inhomogeneous reactive media.  Most practical applications of detonations rarely occur under conditions of perfect homogeneity such as, for example, accident scenarios involving the unintentional release of detonable fuel.  In propulsive applications of detonative combustion, such as the rotating detonation engine (RDE), the detonable mixture is often created by the dynamic injection of fuel and oxidizer immediately ahead of the propagating detonation that may not have time to completely mix, resulting in large spatial variations in chemical reactivity.\cite{Lu2014RDE} Inhomogeneity in density, temperature, and particle velocity might also be present, under certain conditions, due to pre-existing turbulence (again, likely to be encountered in RDEs, for example). Examination of the effect of spatial inhomogeneities on the propagation behavior of gaseous detonation waves is of importance to treat these scenarios.\\

Given such challenging problems for detonation research, the available theoretical tools deriving from first principles are surprisingly simple, perhaps even oversimplified for the task of describing detonation propagation. Most theoretical models of detonation are developed from the assumption of the steady, quasi-one-dimensional Zel'dovich-von Neumann-D\"{o}ring (ZND) solution satisfying a generalized Chapman-Jouguet (CJ) criterion, i.e., a vanishing thermicity at the point in the reaction zone where the flow moves away from the leading shock at the local speed of sound.\cite{Fickett1979,HigginsChapter2} In these models, the reactive medium in which the detonation wave propagates is always considered to be spatially homogeneous based on the averaged thermodynamic, flow, and chemical properties. The question then arises as to whether the propagation behavior of detonation waves that are influenced by the spatial inhomogeneities of the reactive medium (or the inherent spatio-temporal variations that exist in cellular detonations in homogeneous media) can be accurately predicted by these simple models based on an averaging treatment.\\

Answering the above-mentioned question was recently attempted by Mi \textit{et al}. \cite{Mi2016JFM} In their study, the propagation speed of a detonation, resulting from a medium that consists of spatially discretized energy sources separated by regions of inert material, was examined via one-dimensional, direct numerical simulations. In the cases of highly discretized energy sources, the resulting detonation velocity was observed to be greater than the predicted CJ velocity of a homogenized medium with the same amount of energy release (for the case of ratio of a specific heat capacity $\gamma=1.1$, the average wave speed was nearly $15\%$ greater than the CJ speed). These significant deviations from the CJ prediction were hypothesized to be indicative of weak detonations with a non-equilibrium state at the effective sonic surface. In their study, an artificial mechanism of energy deposition, i.e., a discrete source that is instantaneously triggered by the passage of the leading shock, independent of the shock strength, after a prescribed delay time, was implemented due to its simplicity. Hence, a more realistic mechanism of heat release, wherein the energy release evolves from the reactive media itself, depending upon the local thermodynamic state, must be incorporated in further investigations of this problem.\\

\begin{figure}
\centerline{\includegraphics[width=1.0\textwidth]{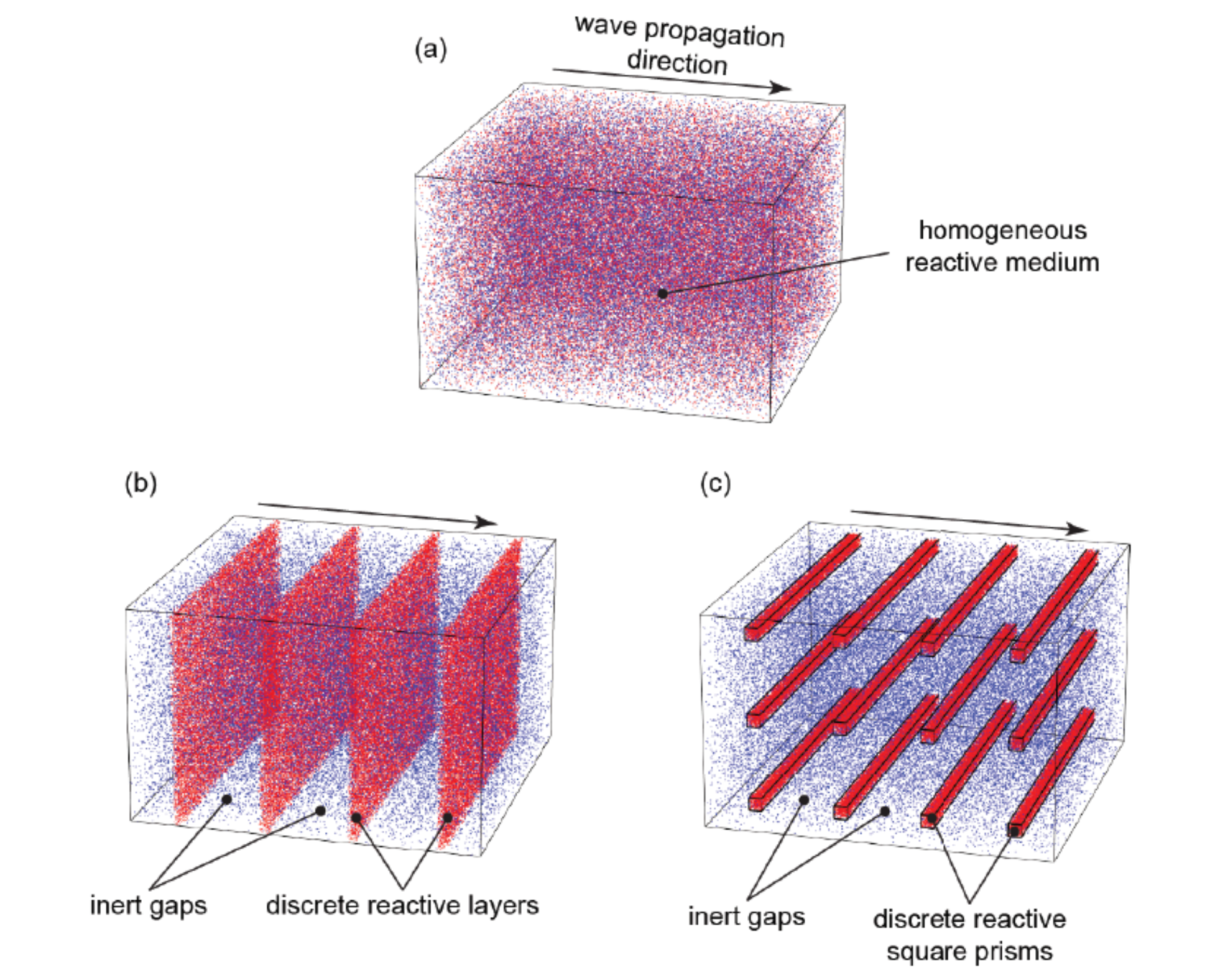}}
		\caption{Conceptual illustrations of a reactive system with (a) energy sources (red dots) homogeneously embedded within an inert medium (blue dots), (b) energy sources collected into spatially discretized layers (or sheets), and (c) energy sources collected into square-based prisms  separated by inert regions.}
	\label{Fig1}
\end{figure}

In the present study, the effect of both one- and two-dimensional spatial inhomogeneities on the propagation speed of gaseous detonation waves without losses is computationally examined. Since a typical detonable mixture of gases is governed by activated chemical reactions, single-step Arrhenius kinetics, as the simplest candidate reaction model, is incorporated into the system. The spatial discretization of energy can be realized, as illustrated in Fig.~\ref{Fig1}(b), via concentrating the reactant into layers (or sheets), standing perpendicular to the direction of detonation wave propagation, separated by regions of inert gas. This arrangement of discrete sources is similar to that used in \cite{Mi2016JFM} and can be implemented in both one- and two-dimensional simulations. Another way to discretize the reactive medium is by concentrating the reactant into infinitely long square-based prisms laying along an axis that is perpendicular to the direction of detonation wave propagation, as shown in Fig.~\ref{Fig1}(c). This arrangement can be implemented in two-dimensional simulations as an array of square sources. These two arrangements of spatial inhomogeneities are referred to as reactive \textit{layers} and \textit{squares}, respectively, in this paper.\\

The first objective of this study is to examine whether the super-CJ wave propagation, which was identified in \cite{Mi2016JFM} for the cases with highly discrete sources, still occurs in a one- or two-dimensional gaseous detonation system with state-dependent Arrhenius kinetics. The simulation results are then analyzed via a spatio-temporal averaging procedure to further elucidate the physical mechanism that is responsible for this super-CJ wave speed. By performing parametric studies, a continuous transition from the continuum CJ propagation to the super-CJ waves in extremely discretized reactive media, i.e., a sequence of point-source blasts that in turn trigger the next source, is systematically explored and analyzed.\\

This paper is organized as follows. In Sec.~\ref{Sec2}, the problem statement and the governing equations of the proposed system are introduced. Section~\ref{Sec3} describes the numerical methodology used to solve the governing equations. The results of sample one- and two-dimensional wave structures, the history of instantaneous propagation speed, and the averaged propagation speed as a function of the model parameters are presented in Sec.~\ref{Sec4}. In Sec.~\ref{Sec5}, the procedures of data analysis are described. The findings based upon the simulation results and the analysis are discussed in Sec.~\ref{Sec6} and summarized in the Conclusions (Sec.~\ref{Sec7}). The detailed derivation of the governing equations based on the averaged properties can be found in the Appendix.\\

\section{Problem Statement}
\label{Sec2}
The detonable mixtures are modelled to be calorically perfect (i.e., with a fixed ratio of specific heats $\gamma$) and have the potential to release chemical energy with a specific heating value $\widetilde{Q}$ ($\mathrm{J}/\mathrm{kg}$). The tilde ``$\sim$'' denotes a dimensional quantity. The flow variables, density, pressure, temperature, and particle velocity ($x$- and $y$-components), are non-dimensionalized with reference to the initial state ahead of the leading shock, i.e., $\rho = \widetilde{\rho}/\widetilde{\rho}_0$, $p = \widetilde{p}/\widetilde{p}_0$, $T = \widetilde{T}/\widetilde{T}_0$, $u = \widetilde{u}/\sqrt{\widetilde{p}_{0}/\widetilde{\rho}_{0}}$, and $v = \widetilde{v}/\sqrt{\widetilde{p}_{0}/\widetilde{\rho}_{0}}$, respectively. The subscript ``$0$'' indicates the pre-shock, initial state of the reactive medium. The properties of a thermodynamic state are related via the ideal gas law, i.e., $\widetilde{p}=\widetilde{\rho}\widetilde{\mathrm{R}}\widetilde{T}$, where $\widetilde{\mathrm{R}}$ is the gas constant, or $p=\rho T$ in dimensionless form. The heat release $\widetilde{Q}$ is non-dimensionalized as $Q=\widetilde{Q}/ \left( \widetilde{p}_{0}/\widetilde{\rho}_{0} \right)$. Applying the CJ criterion, the velocity of a detonation wave propagating in a uniform reactive medium with the heat release $Q$ can be calculated via the following relation,
\begin{equation}
V_{\mathrm{CJ}} = M_{\mathrm{CJ}} c_0 = \sqrt{\frac{\gamma^{2}-1}{\gamma}Q+\sqrt{\left( \frac{\gamma^{2}-1}{\gamma}Q+1 \right)^{2}-1}+1} \cdot \sqrt{\gamma}
\label{Eq1}
\end{equation}
where $M_\mathrm{CJ}$ is the CJ Mach number and $c_0 = \sqrt{\gamma}$ is the non-dimensionalized initial speed of sound. The average propagation speed resulting from each inhomogeneous scenario simulated in this study will be compared with the $V_{\mathrm{CJ}}$ corresponding to a homogeneous reactive system with the same average energy release $Q$.\\

The non-linear, unsteady gasdynamics of the system is described by the two-dimensional (or one-dimensional) reactive Euler equations in the laboratory-fixed reference frame:
\begin{equation}
\frac{\partial \bf{U}}{\partial t} + \frac{\partial \bf{F}\left( \bf{U} \right)}{\partial x} + \frac{\partial \bf{G}\left( \bf{U} \right)}{\partial y} = \bf{S}\left( \bf{U} \right)
\label{Eq2}
\end{equation}
where the conserved variable $\bf{U}$, the convective fluxes $\bf{F}$ and $\bf{G}$, and reactive source term $\bf{S}$ are, respectively,
\begin{equation}
 \bf{U}=
 \begin{pmatrix}
 \rho \\
\rho u \\
\rho v \\
\rho e \\
\rho Z
 \end{pmatrix}
~\qquad \bf{F} =
 \begin{pmatrix}
 \rho u \\
\rho u^2+p \\
\rho uv \\
(\rho e+p)u \\
\rho Zu
 \end{pmatrix}
~\qquad \bf{G} =
 \begin{pmatrix}
 \rho v \\
\rho uv \\
\rho v^2+p \\
(\rho e+p)v \\
\rho Zv
 \end{pmatrix}
~\qquad \bf{S} =
 \begin{pmatrix}
 0 \\
0 \\
0 \\
0 \\
\rho \Omega
 \end{pmatrix}
\label{Eq3}
\end{equation}
In the above equations, $e$ is the non-dimensional specific total energy, and $Z$ is the reaction progress variable, or the normalized concentration of reactant, which varies between $1$ (unreacted) and $0$ (fully reacted). For a homogeneous reactive system, the specific total energy is defined as $e=p/(\gamma-1)\rho+(u^2+v^2)/2+ZQ$. In this study, the reaction rate $\Omega=\partial Z/\partial t$ is governed by single-step Arrhenius chemical kinetics as follows,
\begin{equation}
\Omega = -kZ \times \mathrm{exp}\left(-E_{\mathrm{a}}/T \right)
\label{Eq4}
\end{equation}
where $k$ and $E_\mathrm{a}$ are the dimensionless pre-exponential factor and activation energy, respectively.\\
\begin{figure}
\centerline{\includegraphics[width=0.7\textwidth]{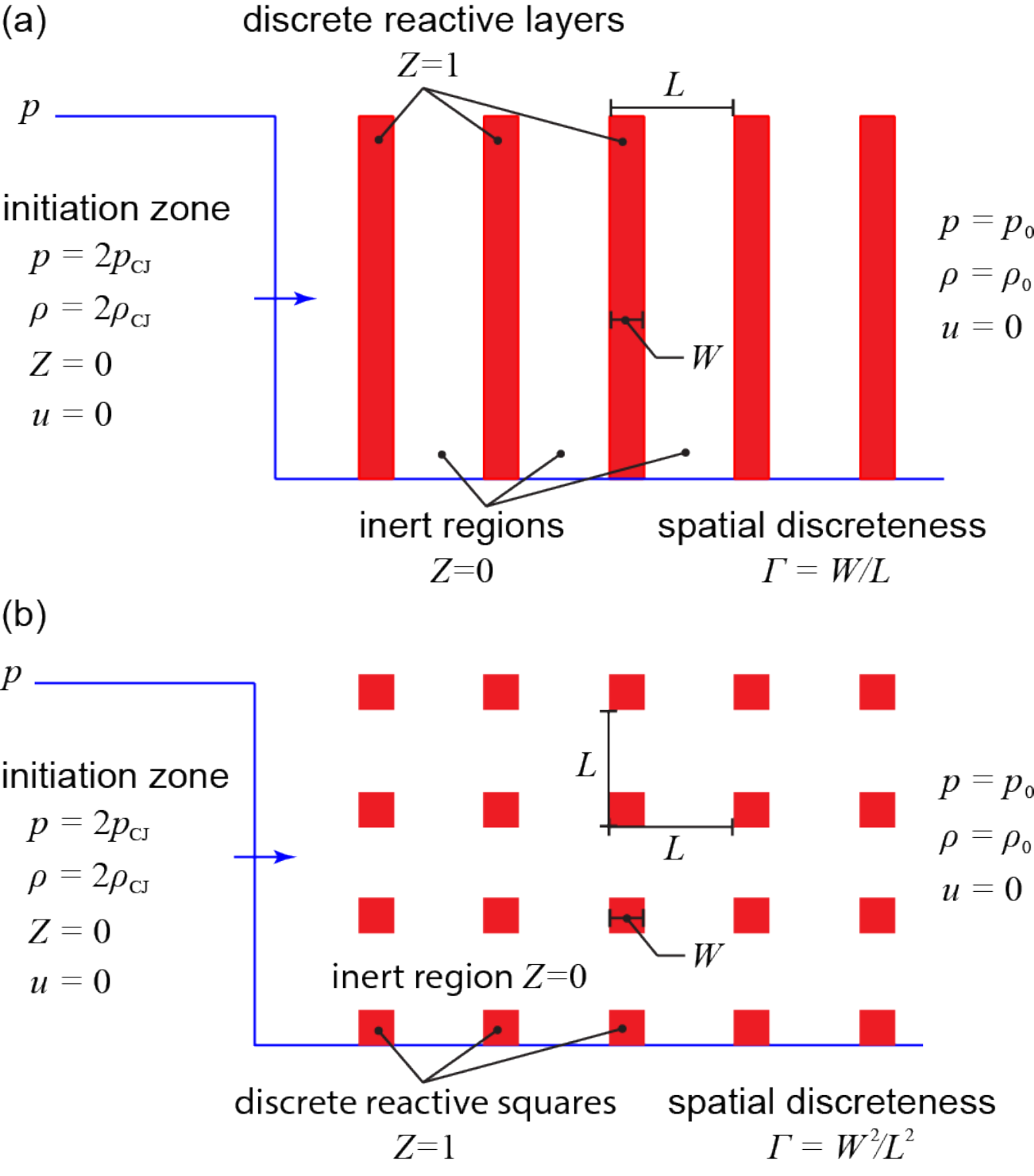}}
		\caption{Schematic showing the initiation method and implementation of spatially discrete reactive (a) layer and (b) squares.}
	\label{Fig2}
\end{figure}

The reactive domain that contains discrete sources is initialized with uniform pressure, density, and particle velocity as $p=1$, $\rho=1$, and $u=0$, respectively. An initiation zone, where pressure and density equal twice the corresponding CJ value, i.e., $p=2p_\mathrm{CJ}$ and $\rho=2\rho_\mathrm{CJ}$, is placed on the left of the reactive domain. A rightward propagating shock wave generated from this initiation zone thus triggers the discrete sources and supports a reaction wave to propagate to the right. The spatial inhomogeneities are introduced into the system as spatially discrete reactive layers or squares separated by inert regions. This spatial discretization is realized by initializing $Z$ as $1$ in the reactive sources and $0$ in the inert regions. As shown in Fig. 2(a), the reactive layers with a width $W$ are distributed in the domain with a regular spacing $L$ between each two consecutive layers. Thus, the initial distribution of $Z$ in space can be described as a summation of regularly spaced, one-dimensional Heaviside (boxcar) functions,
\begin{equation}
Z \left(x,y,t=0 \right) = \sum\nolimits_{i} \mathrm{H} \left( x-iL \right) \mathrm{H} \left( iL+W-x \right)
\label{Eq5}
\end{equation}
where $i$ is the index of each discrete reactive layer. The scenario with discrete reactive squares is shown in Fig. 2 (b). The side length of each square source is $W$ and the spacing between each two neighboring sources is $L$. In this case, the initial distribution of $Z$ can be described as a summation of regularly spaced, two-dimensional Heaviside functions,
\begin{equation}
Z \left(x,y,t=0 \right) = \sum\nolimits_{i} \sum\nolimits_{j} \mathrm{H} \left( x-iL \right) \mathrm{H} \left( iL+W-x \right) \mathrm{H} \left( y-jL \right) \mathrm{H} \left( jL+W-y \right)
\label{Eq6}
\end{equation}
where $i$ and $j$ are the indices of each discrete reactive square in $x$- and $y$-directions, respectively.\\

The spatial discreteness parameter $\Gamma$ is defined as $\Gamma=W/L$ for the cases with reactive layers and $\Gamma=W^2/L^2$ for the cases with reactive squares. In the limit of $\Gamma \to 1$, the initial distribution of $Z$ becomes uniform in the reactive medium; in the limit of $\Gamma \to 0$, the spatially discrete source approaches a $\mathrm{\delta}$-function in space. In order to maintain the overall amount of energy release $Q$ the same as the homogeneous case ($\Gamma=1$), the actual heat release associated with each discrete source must be increased according to the prescribed spatial discreteness $\Gamma$. Hence, for the cases with discrete reactive sources, the specific total energy is formulated as $e=p/(\gamma-1)\rho+(u^2+v^2)/2+ZQ/\Gamma$.\\

This current study is focused on exploring the effect of spatial discreteness $\Gamma$ and source spacing $L$ on the wave propagation behavior in an inhomogeneous reactive medium. The values of $Q$ and $\gamma$ are chosen to be $Q=50$ and $\gamma=1.2$ to represent a typical gaseous detonable mixture. The dimensionless activation energy is chosen to be $E_\mathrm{a}=20$ to ensure stable detonation propagation in the one-dimensional, homogeneous system, which can be used as a control case to more clearly identify the effect of the spatial inhomogeneities and intrinsic multi-dimensional instabilities on the resulting propagation behavior.\cite{LeeStewart1990,Short1998} The pre-exponential factor $k=16.45$ is chosen so that the half-reaction-zone length for the homogeneous case is unity. 

\section{Numerical Methodology}
\label{Sec3}
Two independently-written simulation codes were used to solve the one- and two-dimensional reactive Euler equations. Both of them were based upon a uniform Cartesian grid. The one-dimensional simulation code used the MUSCL-Hancock TVD Godunov-type finite-volume scheme \cite{Toro2009} with an exact Riemann solver and the van Leer non-smooth slope limiter. The reaction process in the one-dimensional simulations was solved using a second-order, two-stage explicit Runge-Kutta method. The Strang splitting method was used in order to maintain second-order accuracy.\cite{Strang1968} The two-dimensional simulation code was also based upon the MUSCL-Hancock scheme but with an HLLC approximate solver for the Riemann problem.\cite{Morgan2013,KiyandaNg2015} This code was implemented in Nvidia's CUDA programming language and performed on a Nvidia Tesla K40M GPU computing processor to accelerate the simulation runs.\\

In each case simulated in this study, the length (i.e., size in the $x$-direction) of the entire domain was at least $50$ times the source spacing $L$. For the cases with reactive squares, transverse width (i.e., size in the $y$-direction) was at least $5L$. In order to have a better algorithmic efficiency, instead of the entire domain, the simulations were only performed in a window that enclosed the leading wave complex at each time step. A window size (in the $x$-direction) of $10L$ (i.e., containing $10$ discrete reactive layers or $10$ vertical arrays of reactive squares) was used and was verified to be sufficient to capture all of the dynamics contributing to the propagation of the leading shock. Once the leading shock front reached the end (right boundary) of the computational window, the window frame (i.e., left and right boundaries) was advanced by half of the window size $5L$. A transmissive boundary condition was applied on both left and right boundaries of the computational window. On the top and bottom boundaries, a periodic boundary condition was applied to simulate a detonation wave propagating in an infinitely wide domain without experiencing any losses due to lateral expansion. The minimum numerical resolution used in this study was $20$ computational cells per half-reaction-zone (unity) length of the homogeneous case, i.e., $\mathrm{\Delta}x=0.05$. For cases with very small source spacing, e.g., $L=1$, a high numerical resolution of $100$ computational cells per half-reaction-zone length was used to ensure a sufficient number ($\sim 10$) of computational cells within a discrete source.\\

\section{Result}
\label{Sec4}
Results from three different cases are presented here:  reactive layers in one dimension, reactive layers in two dimensions, and reactive squares in two dimensions.  For each case, a snapshot of the flow field will be shown, followed by the velocity history. The measured average velocity, $V_\mathrm{avg}$, will be presented as a function of spatial discreteness $\Gamma$ and source spacing $L$. Since $Q=50$, $\gamma = 1.2$, and $E_\mathrm{a} = 20$ are fixed in this current study, only the values of $L$ and $\Gamma$ are mentioned for each specific case of simulation presented in the remainder of this paper. In Figs.~\ref{Fig6} and \ref{Fig9} where the results of $V_\mathrm{avg}$ are presented, the data points for the cases with one-dimensional reactive layers, two-dimensional reactive layers, and two-dimensional reactive squares are plotted as blue circles, green diamonds, and red squares, respectively; solid symbols are for the cases with a fixed $L$ and various $\Gamma$, while open symbols are for the cases with a fixed $\Gamma$ and various $L$.

\subsection{One-dimensional reactive layers}
\label{Sec4_1}
\begin{figure}
\centerline{\includegraphics[width=1.0\textwidth]{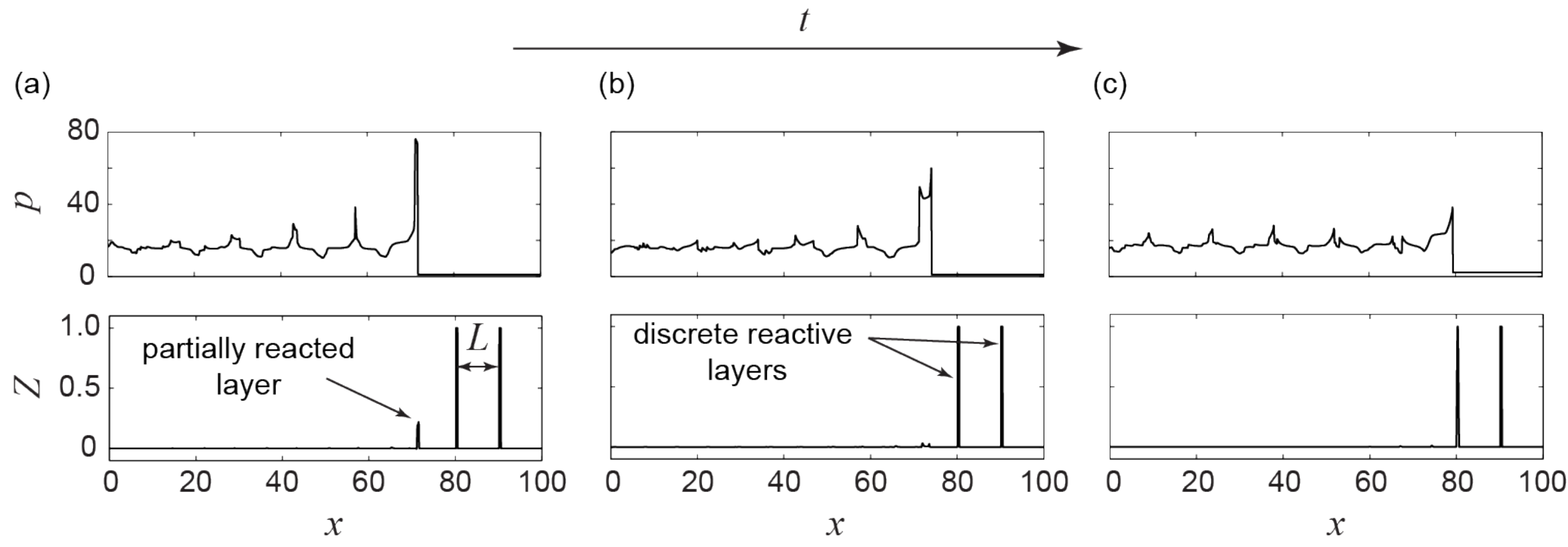}}
		\caption{Time evolution (from (a) to (c)) of the pressure (top row) and reaction progress variable (bottom row) profiles for a one-dimensional simulation with reactive layers and the following parameters: $\Gamma=0.04$ and $L=10$.}
\label{Fig3}
\end{figure}

The sample result plotted in Fig.~\ref{Fig3} shows the time evolution (from (a) to (c)) of the pressure (top row) and reaction progress variable (bottom row) profiles of the computational domain for a simulation with $\Gamma=0.04$ and $L=10$ (spacing between two sources). The leading wave front propagates rightward in this figure. The $\mathrm{\delta}$-function-like, vertical spikes in the profile of $Z$, as indicated in the figure, are the discrete reactive layers where chemical energy is highly concentrated. As shown in Fig.~\ref{Fig3}(a), the peak in the pressure profile is associated with a strong exothermic reaction upon the leading shock encountering one of these reactive layers. The shorter spike in the $Z$ profile in Fig.~\ref{Fig3}(a) corresponds to this partially reacted discrete source shortly after being shocked. As shown in Fig.~\ref{Fig3}(b), forward- and backward-running blast waves that are generated by this strong exothermic reaction can be identified in the pressure profile. Downstream from the leading shock, the pressure profile consists of a large number of decaying and interacting blast waves generated by the earlier sources.
The history of the instantaneous propagation speed $V$ normalized by $V_\mathrm{CJ}$ for the same case is plotted in Fig.~\ref{Fig5}(a) as a function of the leading shock position $x_\mathrm{s}$. The trajectory of the leading shock $x_\mathrm{s}(t)$ can be obtained from the simulation by finding the location where pressure increases to $p=1.5$ from its initial, pre-shock state $p_0=1$ at each time step. The instantaneous propagation speed $V$ can then be calculated by numerically differentiating $x_\mathrm{s}(t)$ over time. After a short process of initiation (over approximately $5$ sources), the wave propagation becomes periodic as shown in the inset in Fig.~\ref{Fig5}(a). A cycle of pulsation in wave velocity, $V$, occurs over a length that is the same as the spacing between two reactive layers, $L$. An averaged propagation speed can be measured over a long distance (about $40$ sources). The average wave speeds $V_\mathrm{avg}$, normalized by $V_\mathrm{CJ}$, resulting from the one-dimensional simulations are plotted as functions of $\Gamma$ and $L$ (as solid and open blue circles) in Fig.~\ref{Fig6}(a) and (b), respectively. As $\Gamma$ decreases from $1$ to $0.01$ or $L$ increases from $1$ to $200$, $V_\mathrm{avg}$ increases from $V_\mathrm{CJ}$ and asymptotically approaches a plateau value that is approximately $9$-$10\%$ greater than $V_\mathrm{CJ}$.

\subsection{Two-dimensional reactive layers}
\label{Sec4_2}
\begin{figure}
\centerline{\includegraphics[width=1.0\textwidth]{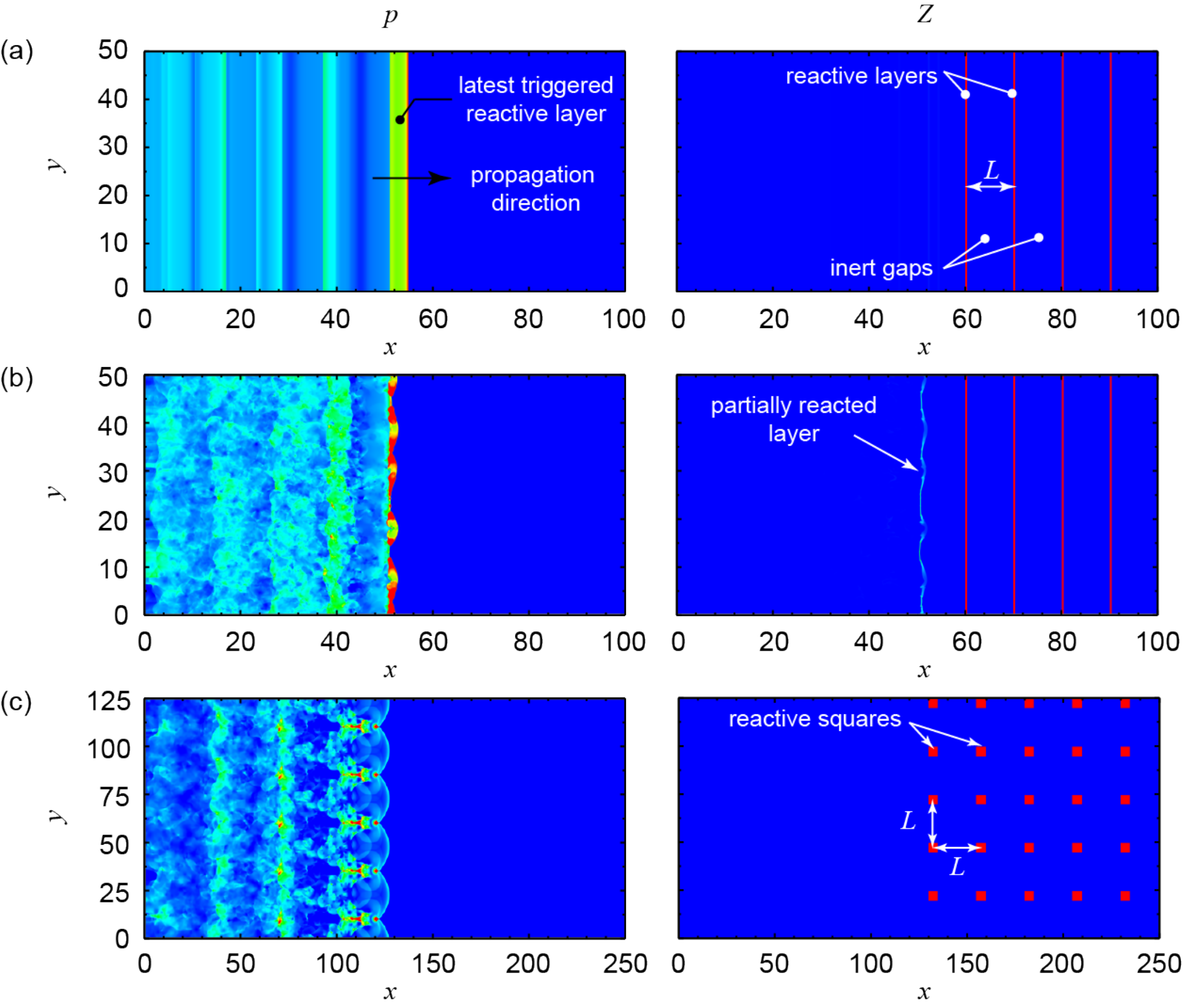}}
		\caption{Sample two-dimensional contours of the pressure (left column) and reaction progress variable (right column) for the case with reactive layers at (a) early ($t=30.2$) and (b) later ($t=140.5$) times with $\Gamma=0.04$ and $L=10$, and (c) for the case with reactive squares, $\Gamma=0.04$, and $L=25$.}
\label{Fig4}
\end{figure}

The sample results plotted in Fig.~\ref{Fig4}(a) and (b) are the two-dimensional contours of the pressure (left column) and reaction progress variable (right column) at early ((a) $t=30.2$) and later (b) $t=140.5$) times for a simulation of discrete reactive layers with $\Gamma=0.04$ and $L=10$. The leading wave front propagates rightward in this figure. The red, vertical lines in the contours of $Z$ are the highly concentrated, reactive layers. At early times, as shown in Fig.~\ref{Fig4}(a), the resulting wave structure remains transversely planar (uniform in the $y$-direction). Forward- and backward-running blast waves associated with high pressure (yellow-red) regions can be clearly identified in Fig.~\ref{Fig4}(a). At later times, as shown in Fig.~\ref{Fig4}(b), significant instabilities have developed, resulting in a no longer planar but highly irregular wave structure. The history of $V/V_\mathrm{CJ}$ as a function of leading shock position $x_\mathrm{s}$ is plotted in Fig.~\ref{Fig5}(b). At each time step, the leading shock position is found along the middle line in $y$-direction of the two-dimensional domain (at $y=25$) using the same technique described in Sec.~\ref{Sec4_1}.\\
As shown in Inset $\mathrm{I}$ of Fig.~\ref{Fig5}(b), $V$ varies in a regularly periodic fashion over a length scale of $L$. After the wave propagates over more than $700$ sources, the variations in $V$ become irregular and exhibit much larger amplitudes. As shown in Inset $\mathrm{II}$, the spacing between two consecutive peaks in $V$ is no longer a constant distance $L$. Note that, before the onset of instabilities, the propagation dynamics resulting from this two-dimensional case with reactive layers are identical to those of the one-dimensional case with the same parameter values (shown in Fig.~\ref{Fig5}(a)). The average propagation velocities reported in this paper were measured over a sufficiently long distance (more than $40$ sources) after the instabilities had fully developed. The $V_\mathrm{avg}$ values resulting from the two-dimensional simulations with reactive layers are plotted as functions of $\Gamma$ and $L$ (as solid and open green diamonds) in Fig.~\ref{Fig6}(a) and (b), respectively. As $\Gamma$ decreases or $L$ increases, $V_\mathrm{avg}$ increases from $V_\mathrm{CJ}$ to super-CJ values. In Fig.~\ref{Fig6}(b), as $L$ increases from $5$ to $200$, $V_\mathrm{avg}$ asymptotically approaches a plateau value that is nearly $10\%$ greater than $V_\mathrm{CJ}$, which is approximately the same as that resulting from the one-dimensional cases. The $V_\mathrm{avg}$ of the two-dimensional simulations are less than those of the corresponding values of the one-dimensional simulations for the same value of $\Gamma$ and $L$.

\subsection{Two-dimensional reactive squares}
\label{Sec4_3}
The sample results plotted in Fig.~\ref{Fig4}(c) are the two-dimensional contours of the pressure (left figure) and reaction progress variable (right figure) at early and later times for a simulation of discrete reactive squares with $\Gamma=0.04$ and $L=25$. The leading wave front propagates rightward in this figure. The red squares in the contour of $Z$ are the highly concentrated sources of energy. As shown in the contour of pressure in Fig.~\ref{Fig4}(c), the transversely regular, wavy leading wave front, which consists of blast waves generated by the energy release of regularly spaced square sources, can be identified. Downstream (to the left) from the leading shock, the wave structure becomes increasingly irregular.
\begin{figure}
\centerline{\includegraphics[width=0.8\textwidth]{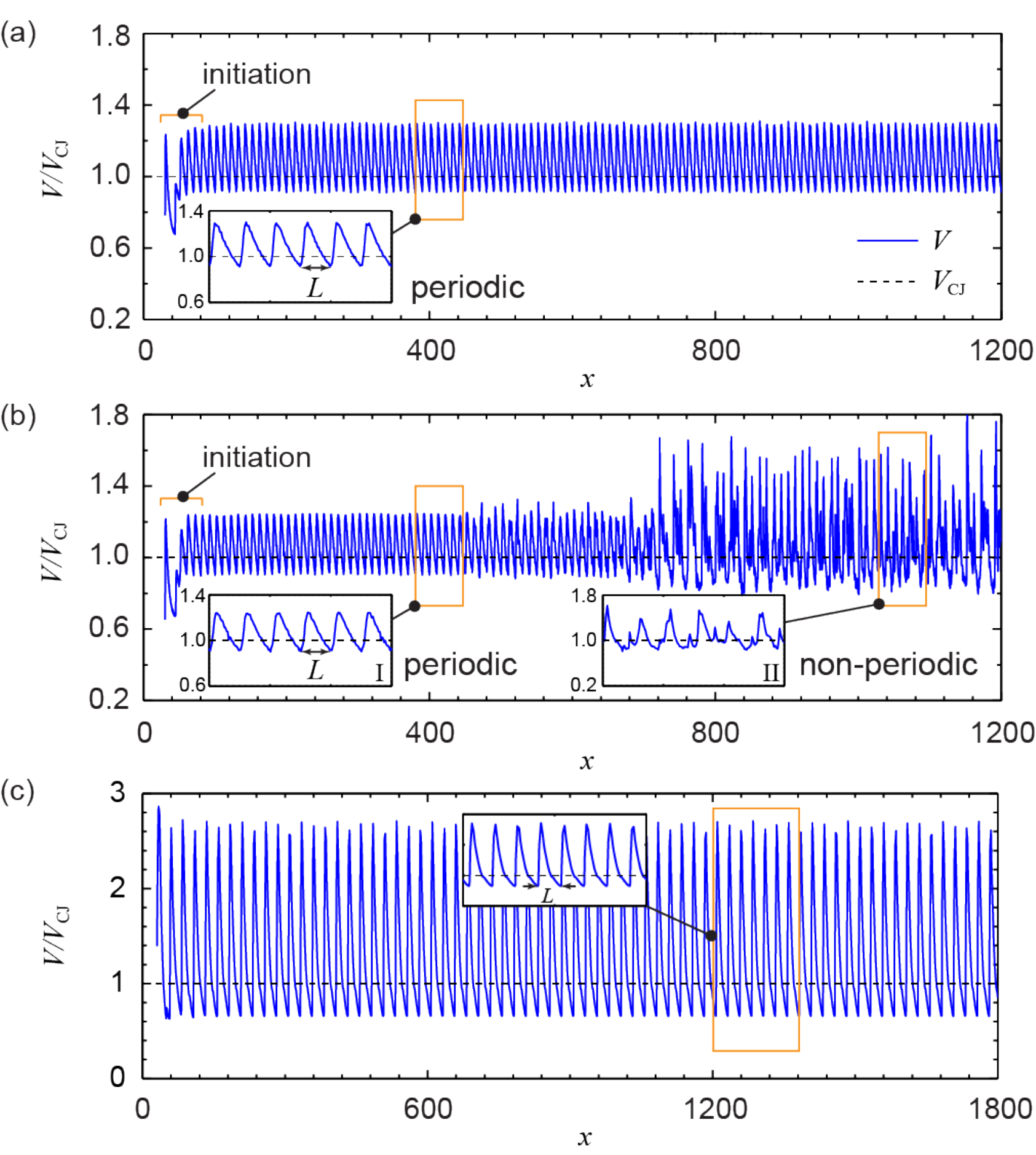}}
		\caption{The history of instantaneous wave propagation velocity normalized by CJ velocity ($V/V_\mathrm{CJ}$) as a function of the leading shock position for the cases with (a) one- and (b) two-dimensional reactive layers ($\Gamma=0.04$ and $L=10$), and (c) reactive squares ($\Gamma=0.04$ and $L=25$).}
\label{Fig5}
\end{figure}
In the plot of $V/V_\mathrm{CJ}$ as a function of leading shock position (Fig.~\ref{Fig4}(c)), a regularly periodic variation in $V$ over a length of $L$, as shown in the inset of Fig.~\ref{Fig5}(c), persists throughout the simulation containing $120$ vertical arrays of square sources. The leading shock position is again defined as that along the middle line in $y$-direction of the two-dimensional domain (at $y=62.5$). The values of $V_\mathrm{avg}$ resulting from the two-dimensional simulations with reactive squares are plotted as functions of $\Gamma$ and $L$ (as solid and open red squares) in Fig.~\ref{Fig6}(a) and (b), respectively. As $\Gamma$ decreases or $L$ increases, $V_\mathrm{avg}$ increases from $V_\mathrm{CJ}$ to super-CJ values. In Fig.~\ref{Fig6}(b), at $L=50$, $V_\mathrm{avg}$ reaches the same plateau value (i.e., nearly $10\%$ greater than $V_\mathrm{CJ}$) as that resulting from both the one- and two-dimensional cases with reactive layers. The $V_\mathrm{avg}$ of the two-dimensional, reactive square cases is fairly close to that of the two-dimensional, reactive layer cases, but lower than that of the one-dimensional cases for the same values of $\Gamma$ and $L$.
\begin{figure}
\centerline{\includegraphics[width=1.0\textwidth]{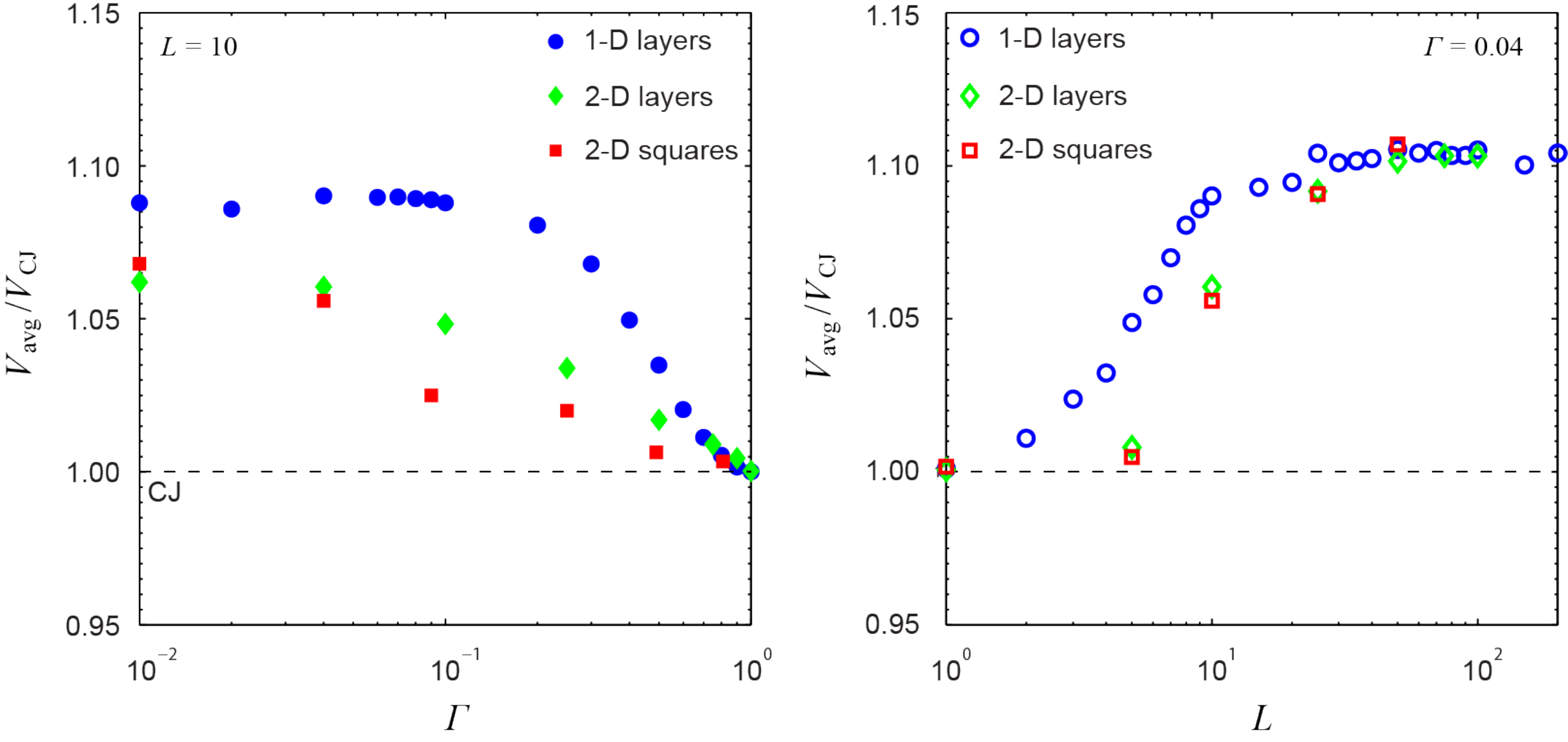}}
		\caption{For cases with one-dimensional reactive layers (blue circles), two-dimensional reactive layers (green diamonds), and two-dimensional reactive squares (red squares), average wave propagation velocity normalized by CJ velocity as (a) a function of $\Gamma$ with $L=10$ and (solid symbols) (b) a function of $L$ with $\Gamma=0.04$ (open symbols).}
	\label{Fig6}
\end{figure}

\section{Analysis}
\label{Sec5}
As shown in Sec.~\ref{Sec4}, a full spectrum of average wave propagation speeds that are significantly greater than $V_\mathrm{CJ}$ is obtained in both one- and two-dimensional systems with discretized energy sources governed by finite-rate, state-dependent Arrhenius kinetics. In order to understand the physical mechanism underlying these super-CJ waves, the simulation results are analyzed in two steps. First, the results of select cases are analyzed via a density-weighted (Favre), spatio-temporal averaging method. Using this analysis, which was introduced to the field of detonation by Lee and Radulescu \cite{LeeRadulescu2005}, Radulescu \textit{et al}. \cite{Radulescu2007JFM}, and Sow \textit{et al}. \cite{Sow2014JFM}. Mi \textit{et al}. interpreted the super-CJ propagation, resulting from a system with highly concentrated sources that instantaneously deposit energy after a fixed delay time, as weak detonations owing to the non-equilibrium condition at the average sonic surface.\cite{Mi2016JFM} The motivation of performing this analysis in the present study is to verify that this mechanism of weak detonation is also responsible for the super-CJ propagation with more realistic reaction kinetics and a higher dimension. Second, with the assistance of an $x$-$t$ diagram constructed from the numerical flow field, a physical parameter, $\tau_\mathrm{c}$, which compares the reaction time of a source $t_\mathrm{r}$ and shock transit time from one source to the next $t_\mathrm{s}$, i.e., $\tau_\mathrm{c} = t_\mathrm{r}/t_\mathrm{s}$, can be determined. This parameter is used to explain the continuous transition of the propagation speed from $V_\mathrm{CJ}$ to the plateau super-CJ value.

\subsection{Averaged steady, one-dimensional wave structure}
\label{Sec5_1}
One- and two-dimensional systems with discrete reactive layers are analyzed using a Favre-averaging approach. The two representative cases selected for further analysis are with $\Gamma=0.04$ and $L=10$. The resulting average wave speed $V_\mathrm{avg}$ in both these one- and two-dimensional cases is approximately $10\%$ greater than the CJ speed. Since the simulations are performed in a lab-fixed reference frame, the data are first transformed into a wave-attached reference frame moving at a constant value of $V_\mathrm{avg}$. For the one-dimensional case, temporal averaging is performed to the transient wave structure as the leading shock propagates over $20$ sources. For the two-dimensional case, the resulting flow field at each time step is first spatially averaged over the transverse ($y$-) direction. The temporal averaging is then performed to the time history of the spatially averaged one-dimensional wave profiles. The two-dimensional results are averaged over the time span required for the leading shock to propagate over $20$ sources. The detailed derivation of the Favre-averaged equations can be found in the Appendix.\\

In Fig.~\ref{Fig7}(a), the averaged pressure $\bar{p}$ for the one-dimensional case is plotted with respect to the wave-attached coordinates $x'$, where $x'=x-V_\mathrm{avg}t$. The sonic point marked as the black circle on the profile of $\bar{p}$ is where the slope of the averaged $u+c$ characteristics equals $0$, i.e., $u^\ast+c^\ast=0$. The average pressure at this averaged sonic point, $\bar{p}_\mathrm{sonic} = 25.1$, is significantly greater than the pressure of the equilibrium CJ state, as indicated by the horizontal dashed line, $p_\mathrm{CJ}=21.5$. This deviation of $\bar{p}_\mathrm{sonic}$ from $p_\mathrm{CJ}$ suggests that equilibrium is not reached as the flow passes through the effective sonic surface. In order to further verify this finding, the thermicity due to the mechanical fluctuation in momentum $\phi_\mathrm{M}$ (blue dash-dot curve), the thermicity due to the thermal fluctuation in total energy $\phi_\mathrm{T}$ (green dash curve), and the exothermicity associated with chemical reaction $\phi_\mathrm{R}$  (red dotted curve) are evaluated and plotted near the average sonic point in Fig.~\ref{Fig7}(b). Thermicity is defined as the terms in the momentum equation that result in a change in the average flow velocity or, equivalently, a change in average pressure of the flow in the reaction zone of a detonation. As shown in this inset, $\phi_\mathrm{M}$, $\phi_\mathrm{T}$, and $\phi_\mathrm{R}$ are still finite, the total thermicity, i.e., $\phi=\phi_\mathrm{M}+\phi_\mathrm{T}+\phi_\mathrm{R}$ (thick black line), reaches zero in the vicinity of the sonic point. These significant fluctuations in momentum and total energy render a non-equilibrium state of the flow upon reaching the effective sonic surface. The derivation of $\phi_\mathrm{M}$, $\phi_\mathrm{T}$, and $\phi_\mathrm{R}$, and the master equation (Eq.~\ref{A13}) that relates the acceleration/deceleration of the averaged flow with the total thermicity and sonic condition are shown in the Appendix.\\
\begin{figure}
\centerline{\includegraphics[width=1.0\textwidth]{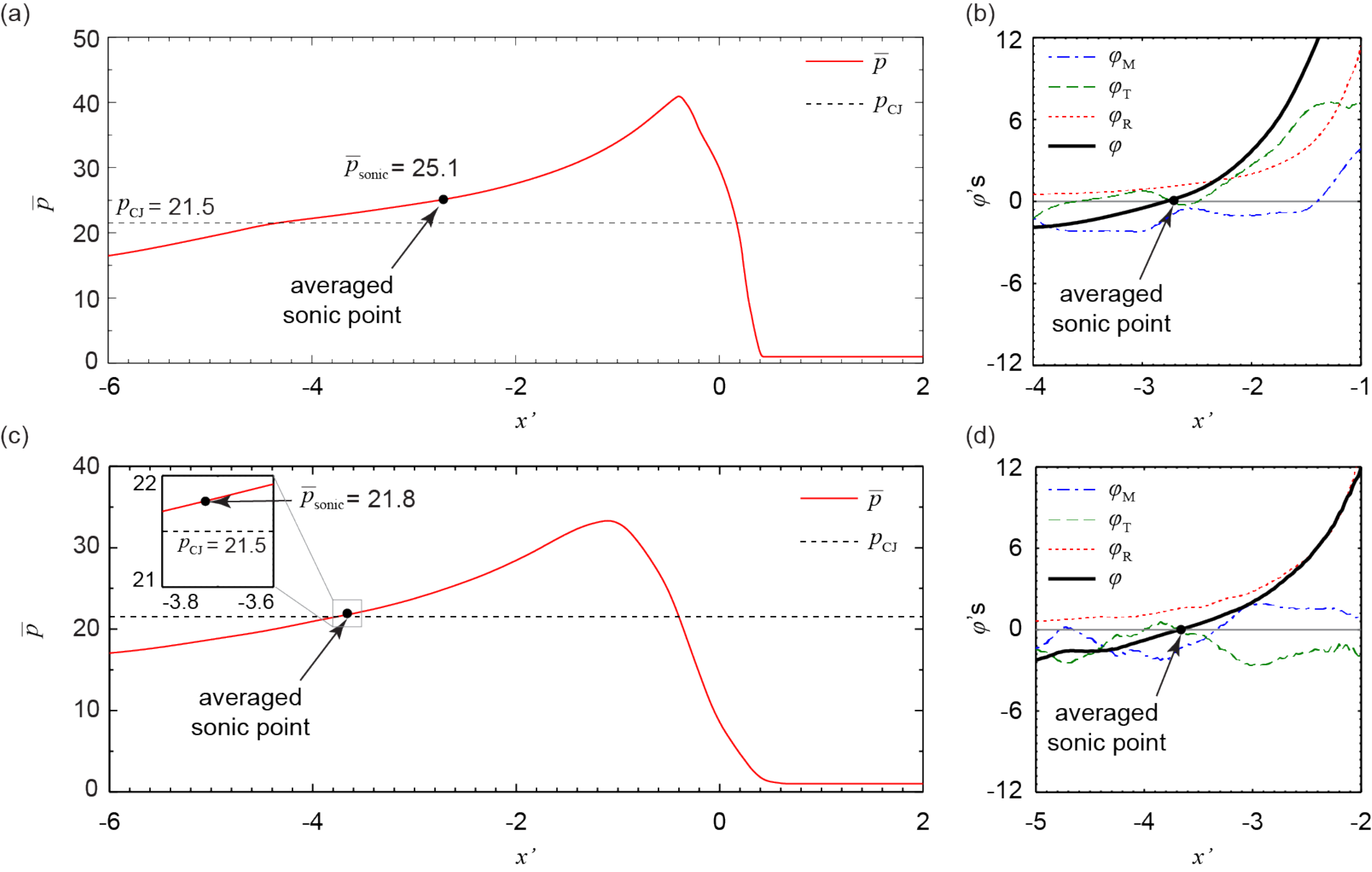}}
		\caption{The spatial profiles of (a) averaged pressure plotted in wave-attached reference frame and (b) the terms of thermicity in the master equation (Eqs.~\ref{A13} and \ref{A14}) plotted in the vicinity of the averaged sonic point for the case with one-dimensional reactive layers, $\Gamma=0.04$, and $L=10$. The spatial profiles of (c) averaged pressure plotted in wave-attached reference frame and (d) the terms of thermicity in the master equation plotted in the vicinity of the averaged sonic point for the case with two-dimensional reactive layers, $\Gamma=0.04$, and $L=10$.}
	\label{Fig7}
\end{figure}
A similar profile of $\bar{p}$ is obtained for the two-dimensional case as shown in Fig.~\ref{Fig7}(c). The jump in pressure associated to the averaged leading shock front is however less sharp (smeared out) than that for the one-dimensional case. As indicated by the black circle in the inset, $\bar{p}_\mathrm{sonic}=21.8$ is close to but still greater than $p_\mathrm{CJ}$ (dashed line). As shown in Fig.~\ref{Fig7}(d), while the exothermic reaction rate still remains significantly positive, and $\phi_\mathrm{M}$ and $\phi_\mathrm{T}$ persist with significantly large amplitudes, the total thermicity $\phi$ vanishes in the immediate vicinity of the average sonic point. Thus, in the two-dimensional case, the non-equilibrium state associated with significant mechanical and thermal fluctuations is identified at the location where the averaged flow encounters the effective sonic surface.

\subsection{Evaluation of $\tau_\mathrm{c}$}
\label{Sec5_2}
As shown in Sec.~\ref{Sec4} (Fig.~\ref{Fig6}), a continuous transition of the propagation speed from $V_\mathrm{CJ}$ to the plateau super-CJ value is found as $\Gamma$ decreases from $1$ to $0$ or $L$ increases. An analogous spectrum of propagation regimes is identified in flame propagation in reactive media with spatially discrete or point-like sources.\cite{GoroshinLeeShoshin1998,Beck2003,Tang2009CTM,Goroshin2011PRE,Tang2011PRE,Mi2016PROCI,Wright2016} A physical parameter, $\tau_\mathrm{c}$, which is the ratio between the heat release time of each source and the characteristic time of heat diffusion between neighboring sources, is used to characterize the corresponding flame propagation regime. Similarly, in this system of discrete source detonations, single-step Arrhenius kinetics with a finite reaction rate permit us to measure the time over which a discrete source (layer or square) releases its chemical energy, $t_\mathrm{r}$. Knowing the trajectory of the leading shock wave, the time required for the wave front to travel from one discrete source to the next, $t_\mathrm{s}$, can also be measured. Thus, the ratio between $t_\mathrm{r}$ and $t_\mathrm{s}$, i.e., $\tau_\mathrm{c}=t_\mathrm{r}/t_\mathrm{s}$, can be evaluated. As the physical significance of $\tau_\mathrm{c}$ related to the wave propagation regimes in discretized reactive media is discussed in Sec.~\ref{Sec6}, this subsection is only focused on presenting an approach to post-processing the simulation data in order to evaluate $\tau_\mathrm{c}$.\\
\begin{figure}
\centerline{\includegraphics[width=1.0\textwidth]{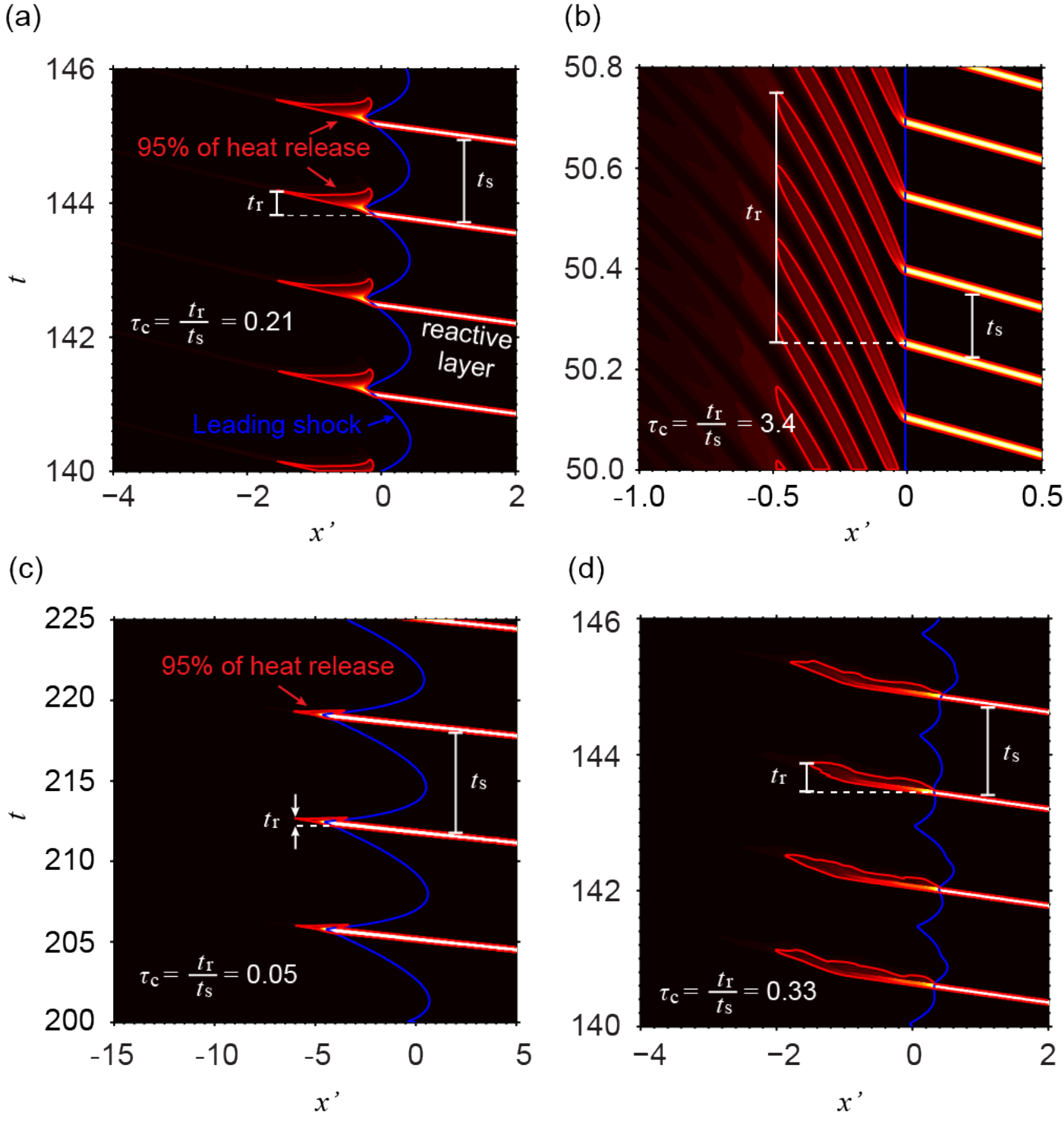}}
		\caption{Contours of reactive progress variable $Z$ plotted in $x'$-$t$ diagrams for the cases with one-dimensional reactive layers with (a) $L=10$, (b) $L=1$, (c) $L=50$, (d) two-dimensional reactive layers with $L=10$, and $\Gamma = 0.04$.}
	\label{Fig8}
\end{figure}

The time evolution of the reaction progress variable $Z$ can be plotted in an $x'$-$t$ diagram where $x'$ is the spatial coordinate in a wave-attached reference frame. This $x'$-$t$ diagram of $Z$ can be directly constructed from the simulation results for the one-dimensional cases. For the two-dimensional simulations, the flow field of $Z$ at each time step first needs to be spatially averaged along the $y$-axis to obtain a one-dimensional profile. Then, the $x'$-$t$ diagram can be constructed based on these averaged profiles of $Z$ from the two-dimensional simulation results. The $x'$-$t$ diagrams of $Z$ for cases with various model parameters are shown in Fig.~\ref{Fig8}.\\
\begin{figure}
\centerline{\includegraphics[width=0.6\textwidth]{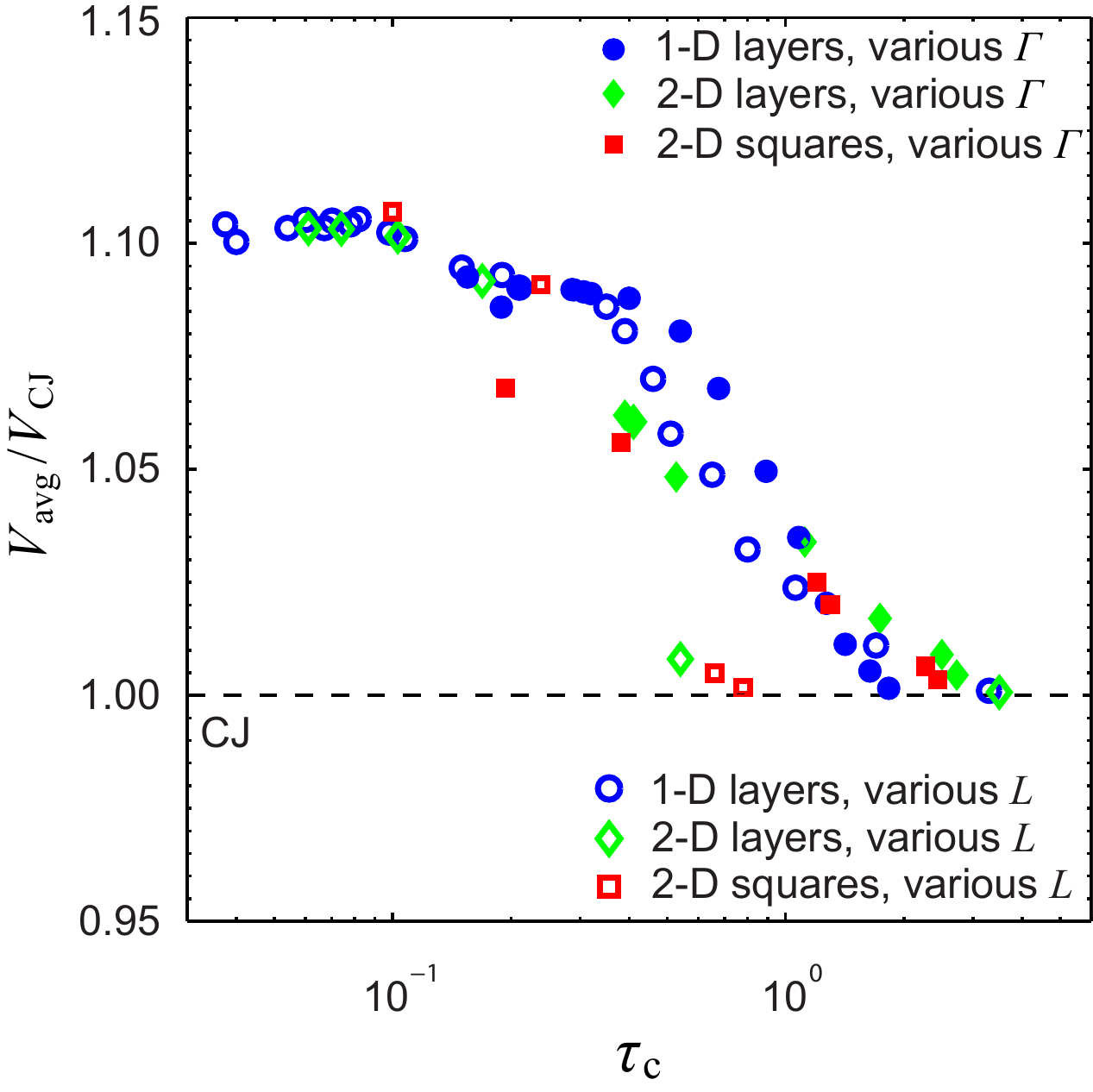}}
		\caption{For cases with one-dimensional reactive layers (blue circles), two-dimensional reactive layers (green diamonds), and two-dimensional reactive squares (red squares), average wave propagation velocity normalized by CJ velocity as a function of $\tau_\mathrm{c}$. The open symbols are for the cases with $\Gamma=0.04$ and various $L$; the solid symbols are for the cases with $L=10$ and various $\Gamma$.}
\label{Fig9}
\end{figure}

Figure~\ref{Fig8}(a), the case with one-dimensional reactive layers ($\Gamma=0.04$ and $L=10$), is taken in this subsection as an example to explain how $\tau_\mathrm{c}$ is determined. The color contour of $Z$ is scaled from bright to dark as $Z=1$ to $0$. Thus, the bright stripes on the right of this figure are the loci of unreacted discrete layers moving (leftwards) towards the leading shock whose trajectory is plotted as the blue curve. The shock transit time between discrete sources $t_\mathrm{s}$ can be obtained by measuring the vertical spacing between two bright stripes. The dark zones separating the discrete layers are the inert regions. The areas of gradual color change emanating from where the leading shock encounters the reactive layers indicate the energy release. The areas of energy release are bounded by a thin red outline, which is the iso-contour of $Z=0.05$, indicating that $95\%$ of the chemical energy initially stored in each source is released within the bounded zone. The reaction time $t_\mathrm{r}$ of each discrete source can be measured as the vertical spacing between the locus of shock-source intersection and the upper bound (in time) of the $95\%$ heat release zone. Although this technique of determining $t_\mathrm{r}$ is somewhat arbitrary, it should be sufficient to characterize the wave propagation regimes as long as this measurement is consistently performed in this study. The results of average propagation velocity normalized by the CJ value for the scenarios of one-dimensional reactive layers (blue circles), two-dimensional reactive layers (green diamonds), and two-dimensional reactive squares (red squares) with various $\Gamma$ (solid symbols) and $L$ (open symbols) can thus be plotted as a function of $\tau_\mathrm{c}$ as shown in Fig.~\ref{Fig9}.

\section{Discussion}
\label{Sec6}
The results presented in this paper show that, in an adiabatic system of discretized energy sources governed by single-step Arrhenius kinetics, waves can propagate, in a self-sustained manner, at a speed that is significantly greater than the CJ value of a homogeneous system with the same amount of overall heat release and without the support of a piston. Based on the analysis presented in Sec.~\ref{Sec5_1} for selected one- and two-dimensional cases, this nearly $10\%$ super-CJ wave propagation can be interpreted as a weak detonation where the flow remains in a non-equilibrium state upon reaching the effective sonic surface. Note that, of all detonation solutions satisfying the conservation laws, the CJ solution with a complete equilibrium state at the sonic surface corresponds to the slowest possible wave speed. By evaluating the terms which comprise the total thermicity in the master equation (Eq.~\ref{A13}) based on the Favre-averaged properties, the non-equilibrium condition at the sonic point is attributed to the intense fluctuations in momentum and total energy of the flow. The generalized-CJ condition, i.e., a vanishing thermicity ($\phi=0$) at the average sonic point ($u^\ast+c^\ast = 0$), is satisfied owing to the balance between the exothermic chemical reaction and the mechanical and thermal fluctuations. The finding of this study incorporating a more realistic, state-dependent reaction model complements the previous study by Mi \textit{et al}.\cite{Mi2016JFM}, verifying that the classic CJ criterion assuming a homogeneous medium based on averaged properties is not always applicable to predict the wave propagation speed in a spatially inhomogeneous system, and further suggesting that the resulting super-CJ propagation is independent of the particular energy deposition mechanism.\\

In the previous work of Mi \textit{et al}. \cite{Mi2016JFM}, where an instantaneous, state- and shock strength-independent mechanism of energy deposition was considered, the spatial coordinate can be normalized by the regular spacing between two consecutive sources. In other words, source spacing $L$ does not affect the wave propagation behavior. In that study, the ratio of specific heat capacity $\gamma$ and the spatial discreteness parameter $\Gamma$ are the only two factors determining the deviation of average wave speed away from the CJ solution. In the current study, however, as a finite-rate, state-dependent reaction rate is incorporated, an additional length scale, i.e., the reaction zone length of a detonation in the homogenized system, comes into play. This physical length scale is a function of $E_\mathrm{a}$, $Q$, and $\gamma$, but independent of source spacing. The source spacing relative to the intrinsic reaction zone length therefore affects the resulting wave propagation behavior.\\

The effect of $L$ on the average wave speed can be identified in Fig.~\ref{Fig6}(b). For a fixed spatial discreteness $\Gamma=0.04$, $V_\mathrm{avg}$ increases from $V_\mathrm{CJ}$ to a plateau value that is $10\%$ greater than $V_\mathrm{CJ}$ as $L$ increases from $1$ (i.e., source spacing equals half-reaction-zone length) to $200$. As $\Gamma$ decreases from $1$ to the limit of $\Gamma \to 0$, a similar trend of $V_\mathrm{avg}$ increasing from the CJ speed to the same plateau value is shown in Fig.~\ref{Fig6}(a). These two asymptotic limits of $\Gamma$ can be understood as follows: When $\Gamma=1$, the source size equals the source spacing; the system is thus continuous, resulting in a CJ propagation speed. As $\Gamma \to 0$, the discrete sources tend to be spatial $\mathrm{\delta}$-functions and release energy nearly instantaneously. In this limit, each source generates forward- and backward-running blast waves. The forward running blast triggers the next source, so the wave propagates via a mechanism of sequentially initiated blast waves by the point sources, which can be qualitatively captured by the heuristic model based on the construction of point-source blast solutions in the Appendix of Ref.~\cite{Mi2016JFM}. Since the variation of $V_\mathrm{avg}$ as a function of $L$ is between the same asymptotic limits as those of $\Gamma$, the underlying mechanisms at these limits must have an equivalent effect on the wave propagation. When $L$ is small, i.e., on the order of the intrinsic half-reaction-zone length, these spatial inhomogeneities are too fine so that the reactive medium is effectively homogenized. In the other limit, where $L$ is hundreds of times larger than the half-reaction-zone length, the time of a discrete source being processed by the leading shock and releasing energy is much shorter than the time required for the leading shock to travel from a source to the next. Hence, given the large time scale of wave propagation, the energy of one source is released effectively instantaneously, and the overall picture of this wave propagation reverts to the case of sequentially triggered point blasts. Note that, since neither losses nor a chemical kinetic cutoff are considered in this system, further increasing $L$ will not qualitatively alter the resulting wave dynamics or lead to quenching.\\

The continuous spectrum of the wave solutions from the effectively homogeneous CJ propagation to a sequence of point-source blasts can be rationalized with the assistance of $\tau_\mathrm{c}$ evaluated via the method presented in Sec.~\ref{Sec5_2}. In other words, the effect of $L$ and $\Gamma$ on the wave propagation speed can be reconciled by considering the $\tau_\mathrm{c}$ parameter. The $x'$-$t$ diagram of $Z$-contour shown in Fig.~\ref{Fig8}(b) is for the case of one-dimensional reactive layers with $\Gamma=0.04$ and $L=1$, where the reaction time $t_\mathrm{r}$ of a source is much longer than the shock transit time $t_\mathrm{s}$, i.e., $\tau_\mathrm{c}$ is significantly greater than unity. This case corresponds to the scenario wherein the very small scale discrete sources are effectively homogenized, and results in a CJ wave speed. Keeping $\Gamma$ fixed at $0.04$ and increasing $L$ to $10$ (for the one-dimensional case), as shown in Fig.~\ref{Fig8}(a), $t_\mathrm{r}$ is still finite but smaller than $t_\mathrm{s}$. In this case, where $\tau_\mathrm{c}=0.21$, $V_\mathrm{avg}$ reaches an intermediate value that is approximately $8.5\%$ greater than $V_\mathrm{CJ}$, but still less than the $10\%$ super-CJ plateau value. For the one-dimensional case with  $\Gamma=0.04$ and $L$ increased to $50$,  as shown in Fig.~\ref{Fig8}(c), $t_\mathrm{r}$ is significantly smaller than $t_\mathrm{s}$, i.e., $\tau_\mathrm{c}=0.05$. The wave propagation in this case is thus via the mechanism of sequentially triggered point-source blasts, and a plateau super-CJ speed is observed.\\

As shown in Fig.~\ref{Fig6}, the resulting $V_\mathrm{avg}$ values in the one- and two-dimensional cases with reactive layers coincide at the CJ and plateau super-CJ limits, but differ over the transitional range of $\Gamma$ and $L$. Over this range, the $V_\mathrm{avg}$ values of the two-dimensional cases are smaller than that of the one-dimensional cases. This difference is due to the fact that, while the detonation in the one-dimensional homogeneous system is stable for the selected parameters ($Q=50$, $\gamma=1.2$, and $E_\mathrm{a}=20$) \cite{LeeStewart1990}, it is intrinsically unstable in a homogeneous two-dimensional system.\cite{Short1998} In addition, the stability analysis which indicates that this system should be stable in one-dimension only applies to homogeneous media.  As the source energy is concentrated into reactive layers or squares, the local heat release increases by a factor of $1/\Gamma$, likely promoting the development of instability. For the cases with large $\Gamma$ and small $L$, which are not severely inhomogeneous, the intrinsic detonation instabilities are likely developed in a two-dimensional system. As shown in the two-dimensional sample result in Fig.~\ref{Fig4}(b), after the instabilities have fully developed, the leading shock front becomes transversely wavy, and thus processes different parts of the discrete reactive layer at different times and with different strength. The spatially smeared shock front in the $\bar{p}$ profile for the two-dimensional case shown in Fig.~\ref{Fig7}(c) is a result of these developed instabilities. Hence, the heat release of a discrete layer is also temporally and spatially smeared out, having a homogenizing effect on the energy deposition. This effect can be verified in the $x'$-$t$ diagram of $Z$-contour for the two-dimensional case with $\Gamma=0.04$ and $L=10$ based on the spatially averaged one-dimensional wave profiles (Fig.~\ref{Fig8}(d)). The $\tau_\mathrm{c}$ for this case is determined as $0.33$, which is greater than that for the one-dimensional case with the same $\Gamma$ and $L$, i.e., $\tau_\mathrm{c}=0.21$, as shown in Fig.~\ref{Fig8}(a) and (d). Correspondingly, the $V_\mathrm{avg}$ resulting from the above-mentioned two-dimensional case is $6.1\%$ greater than $V_\mathrm{CJ}$ while that for the one-dimensional case is $8.5\%$ greater than $V_\mathrm{CJ}$. Therefore, as an alternative to $\Gamma$ and $L$, $\tau_\mathrm{c}$ can be used as a general parameter that quantifies the effect of energy discretization on the wave propagation speed. As demonstrated in Fig.~\ref{Fig9}, the results $V_\mathrm{avg}$ for both one- and two-dimensional cases with various $\Gamma$ and $L$ follow qualitatively the same trend when plotted as a function of $\tau_\mathrm{c}$.\\

In this study, the super-CJ wave propagation is identified in the cases with a two-dimensional arrangement of reactive squares. The super-CJ plateau value and the dependence of the deviation from the CJ propagation on the spatial discreteness $\Gamma$ and spacing between reactive squares $L$ is qualitatively the same as that for the cases with reactive layers. This result suggests that the super-CJ wave propagation and its underlying mechanism due to the spatial inhomogeneities are unlikely an artifact only arising from a one-dimensional system or a system with a one-dimensional, laminar-like arrangement of discrete sources (i.e., reactive layers), but a rather fundamental consequence of multi-dimensionally distributed inhomogeneities on the propagation of reaction waves.\\

Although this study considers simplified scenarios of spatial inhomogeneities, it may capture some details of a detonation propagating in the combustion chamber of a rotating detonation engine with discretely located fuel/oxidizer injection. The scenario with reactive layers resembles the RDE design where detonable gases are axially injected into the annular combustion chamber such as those studied in Refs.~\cite{Hishida2009,Fujii2016,Yao2016}; the RDEs with impinging injection of non-premixed fuel and oxidizer \cite{Kindracki2011,Lin2015,Rankin2016} can be conceptualized as the scenario with discrete reactive squares. The key finding of this current work may explain the $5\%$ super-CJ detonation velocity recently reported by Fujii \textit{et al.}\cite{Fujii2016} for the numerical simulations of a detonation wave propagating in a RDE combustion chamber with relatively widely spaced, premixed gas injection.\\

Drawing inspiration from Vasil'ev and Nikolaev's heuristic model \cite{Vasiliev1978}, which utilized interacting point-source blast waves to mimic detonation cell structure, the two-dimensional arrangement of highly concentrated, reactive squares considered in this study can be used to investigate the wave dynamics of cellular detonations in future efforts. By selecting a source spacing $L$ that is similar to typical detonation cell sizes, the wave structure induced by imposing spatial inhomogeneities can be hypothesized to have a similar effect on the overall propagation behavior and critical limits as those resulting from the intrinsic cellular structure. Spatially regular and random distributions of inhomogeneities can potentially be used to induce wave structures similar to that in weakly and strongly unstable mixtures, respectively.\\

Further development of this detonation system with spatial inhomogeneities will also be carried out by incorporating a multi-step, chain-branching reaction scheme that provides a kinetic quenching mechanism \cite{ShortQuirk1997,NgLee2003}, i.e., a critical temperature below which the exothermic reaction rate is decreased significantly (or quenches). With such a system, it would be possible to examine critical detonation phenomena, for example, a propagation limit in source spacing $L$ beyond which the blast wave generated by a discrete source decays to a shock that is too weak to trigger the exothermic reaction of the subsequent sources.

\section{Conclusion}
\label{Sec7}
The effect of spatial inhomogeneity in the reaction progress variable upon detonation propagation, while maintaining the overall energy release of the medium as constant, has been studied via numerical simulations in one-dimensional systems and in two-dimensional systems of reactive layers and squares governed by activated, Arrhenius kinetics.  The average wave speeds are observed to agree with the predictions of the classical Chapman-Jouguet criterion provided that the time scale of the energy release is greater than the time required for the leading shock to propagate between sources.  This regime is observed if the medium is nearly homogeneous (i.e., with the gaps of inert media being smaller than the reactive areas) or when the spacing between the reactive layers is small in comparison to the half reaction zone length of a detonation in the equivalent homogeneous media.  In sufficiently inhomogeneous media, wherein the spacing between reactive regions is greater than the inherent reaction zone length, average wave propagation speeds significantly greater than the CJ velocity of the equivalent homogenous medium are observed (up to $10\%$).  Based on spatial and temporal averaging of the numerical results, the super-CJ waves can be interpreted as weak detonations wherein the generalized CJ condition applies at a state of non-equilibrium existing at an effective sonic point inside the wave structure, rather than at an equilibrium point located at the end of the reaction zone in the classical CJ detonation criterion.  The non-equilibrium condition in the flow is attributed to persistent fluctuations in momentum and total energy resulting from the intense shock waves generated by the concentrated pockets of energy release.

\appendix*
\section{}
\label{App}
The complete derivation of the Favre-averaged (i.e., density-weighted, spatio-temporally averaged) equations, the master equation, and the thermicity terms ($\phi$, $\phi_\mathrm{M}$, $\phi_\mathrm{T}$, and $\phi_\mathrm{R}$) are presented in this Appendix. The averaging is performed in a reference frame moving at the averaged wave propagation velocity $V_\mathrm{avg}$. In this moving reference frame, the spatial coordinate and the $x$-component of particle velocity are transformed as $x'=x-V_\mathrm{avg}t$ and $u'=u-V_\mathrm{avg}$, respectively. For convenience, $u$ denotes the $x$-component of particle velocity with respect to the moving frame in this Appendix.

A simple spatio-temporal averaging (or only temporal for one-dimensional cases), i.e., Reynolds averaging procedure, is then applied to density and pressure as follows
\begin{equation}
\bar{\rho}\left( x' \right) = \frac{1}{t_{2}-t_{1}} \int_{t_{1}}^{t_{2}} \frac{1}{y_{2}-y_{1}} \int_{y_{1}}^{y_{2}} \rho \left( x',t \right) \mathrm{d}y\mathrm{d}t \;\;\; \mathrm{and} \;\;\;  \rho = \bar{\rho}+\rho^{\circ}
\label{EqA1}
\end{equation}
\begin{equation}
\bar{p}\left( x' \right) = \frac{1}{t_{2}-t_{1}} \int_{t_{1}}^{t_{2}} \frac{1}{y_{2}-y_{1}} \int_{y_{1}}^{y_{2}} p \left( x',t \right) \mathrm{d}y\mathrm{d}t \;\;\; \mathrm{and} \;\;\;  p = \bar{p}+p^{\circ}
\label{A2}
\end{equation}
where $t_1$ and $t_2$ indicate the starting and ending time of the period, and $y_1$ and $y_2$ indicate the lower and upper boundaries of the computational domain in the $y$-direction, over which $\rho$ and $p$ are averaged. The bar ``$\overline{~~}$'' and superscript ``$\circ$'' indicate spatio-temporally averaged variables and their corresponding fluctuating quantities. Favre averaging (i.e., density-weighted averaging) is applied to the particle velocity and reaction progress variable as follows,
\begin{equation}
u^\ast = \frac{\overline{\rho u}}{\bar{\rho}} \;\; \mathrm{and} \;\; u = u^\ast+u''
\label{A3}
\end{equation}
\begin{equation}
Z^\ast = \frac{\overline{\rho Z}}{\bar{\rho}} \;\; \mathrm{and} \;\; Z = Z^\ast+Z''
\label{A4}
\end{equation}
where superscripts ``$\ast$'' and ``$''$'' indicate Favre-averaged variables and their corresponding fluctuating quantities, respectively. The average structure of the wave is therefore governed by the one-dimensional, stationary Favre-averaged Euler equations as follows, 
\begin{equation}
\frac{\partial}{\partial x'} \left( \bar{\rho} u^\ast \right) = 0
\label{A5}
\end{equation}
\begin{equation}
\frac{\partial}{\partial x'} \left( \bar{\rho} {u^\ast}^2 + \bar{p} + \overline{\rho {u''}^2}\right) = 0
\label{A6}
\end{equation}
\begin{equation}
\frac{\partial}{\partial x'} \left( \bar{\rho} e^\ast u^\ast + \bar{\rho}\left(e'' u'' \right)^\ast + \overline{pu} \right) = 0
\label{A7}
\end{equation}
where the averaged specific total energy $e^\ast$ can be expressed as follows,
\begin{equation}
e^\ast = \frac{\bar{p}}{\bar{\rho}(\gamma-1)} + \frac{{u^\ast}^2}{2} +\frac{Z^\ast Q}{\Gamma}
\label{A8}
\end{equation}
Knowing the upstream boundary condition, i.e., the initial state of the region ahead of the leading shock, Eqs.~\ref{A5}-\ref{A7} can be integrated to obtain the following equations,
\begin{equation}
\bar{\rho} u^\ast = V_\mathrm{avg}
\label{A9}
\end{equation}
\begin{equation}
\frac{V_\mathrm{avg}^2}{\bar{\rho}} + \bar{p} + f = V_\mathrm{avg}^2 + 1
\label{A10}
\end{equation}
\begin{equation}
\frac{\gamma \bar{p}}{(\gamma-1) \bar{\rho}} + \frac{{u^\ast}^2}{2} + \frac{Z^\ast Q}{\Gamma} +\frac{g}{V_\mathrm{avg}} = \frac{\gamma}{\gamma-1} + Q + \frac{{V_\mathrm{avg}}^2}{2}
\label{A11}
\end{equation}
where $f = \overline{\rho {u''}^2}$ and $g = \overline{\rho {e''}{u''}} + \overline{p^{\circ}{u''}}$ are the intensities of mechanical and thermal fluctuations, respectively. With the averaged quantities $V_\mathrm{avg}$, $\bar{p}$, $\bar{\rho}$, $u^\ast$, and $Z^\ast$ calculated, $f$ and $g$ can be then evaluated using Eqs.~\ref{A10} and \ref{A11}.\\

The average sound speed, which is assumed to be independent of the intensity of fluctuation, can be calculated as
\begin{equation}
c^\ast = \sqrt{\frac{\gamma \bar{p}}{\bar{\rho}}}
\label{A12}
\end{equation}
The effective sonic point in the one-dimensional averaged wave structure is located at the position at where $u^\ast + c^\ast = 0$. Considering Eqs.~\ref{A5} and \ref{A6} and taking the expression for $e^\ast$ (Eq.~\ref{A8}) into Eq.~\ref{A7}, after some algebraic manipulation, one obtains the so-called master equation as follows,
\begin{equation}
\frac{\mathrm{d}u^\ast}{\mathrm{d}x'} = \frac{\gamma u^\ast \frac{\mathrm{d}f}{\mathrm{d}x'} - (\gamma-1)\frac{\mathrm{d}g}{\mathrm{d}x'} - \frac{(\gamma-1) Q V_\mathrm{avg}}{\Gamma} \frac{\mathrm{d}Z^\ast}{\mathrm{d}x'}}{\bar{\rho} \left( {c^\ast}^2 - {u^\ast}^2 \right)} = \frac{\phi}{\bar{\rho}\left( {c^\ast}^2 - {u^\ast}^2 \right)}
\label{A13}
\end{equation}
where
\begin{equation} 
\begin{split}
\phi_\mathrm{M} & = \gamma u^\ast \frac{\mathrm{d}f}{\mathrm{d}x'} \\
\phi_\mathrm{T} & = -(\gamma-1)\frac{\mathrm{d}g}{\mathrm{d}x'} \\
\phi_\mathrm{R} & = - \frac{(\gamma-1) Q V_\mathrm{avg}}{\Gamma} \frac{\mathrm{d}Z^\ast}{\mathrm{d}x'} \\
\phi & =  \phi_\mathrm{M}+\phi_\mathrm{T}+\phi_\mathrm{R}
\end{split}
\label{A14}
\end{equation}
The master equation describes how the particle velocity of a fluid element traversing through a one-dimensional, steady Favre-averaged wave structure is influenced by the thermicity ($\phi$) due to mechanical fluctuations ($\phi_\mathrm{M}$), thermal fluctuations ($\phi_\mathrm{M}$), and chemical reaction progress ($\phi_\mathrm{R}$). Upon the flow passing through the averaged sonic point where the denominator of the master equation (Eq.~\ref{A13}) equals zero, i.e., ${c^\ast}^2 - {u^\ast}^2 = 0$, the thermicity $\phi$ must vanish, i.e., $\phi = 0$. Otherwise, a singularity would be encountered at the averaged sonic point. Thus, the condition $\phi = 0$ at the sonic surface that permits a singularity-free wave structure is known as the \textit{generalized CJ condition}.

\bibliographystyle{apsrev4-1}
\bibliography{detonation}

\begin{thebibliography}{35}%
\makeatletter
\providecommand \@ifxundefined [1]{%
 \@ifx{#1\undefined}
}%
\providecommand \@ifnum [1]{%
 \ifnum #1\expandafter \@firstoftwo
 \else \expandafter \@secondoftwo
 \fi
}%
\providecommand \@ifx [1]{%
 \ifx #1\expandafter \@firstoftwo
 \else \expandafter \@secondoftwo
 \fi
}%
\providecommand \natexlab [1]{#1}%
\providecommand \enquote  [1]{``#1''}%
\providecommand \bibnamefont  [1]{#1}%
\providecommand \bibfnamefont [1]{#1}%
\providecommand \citenamefont [1]{#1}%
\providecommand \href@noop [0]{\@secondoftwo}%
\providecommand \href [0]{\begingroup \@sanitize@url \@href}%
\providecommand \@href[1]{\@@startlink{#1}\@@href}%
\providecommand \@@href[1]{\endgroup#1\@@endlink}%
\providecommand \@sanitize@url [0]{\catcode `\\12\catcode `\$12\catcode
  `\&12\catcode `\#12\catcode `\^12\catcode `\_12\catcode `\%12\relax}%
\providecommand \@@startlink[1]{}%
\providecommand \@@endlink[0]{}%
\providecommand \url  [0]{\begingroup\@sanitize@url \@url }%
\providecommand \@url [1]{\endgroup\@href {#1}{\urlprefix }}%
\providecommand \urlprefix  [0]{URL }%
\providecommand \Eprint [0]{\href }%
\providecommand \doibase [0]{http://dx.doi.org/}%
\providecommand \selectlanguage [0]{\@gobble}%
\providecommand \bibinfo  [0]{\@secondoftwo}%
\providecommand \bibfield  [0]{\@secondoftwo}%
\providecommand \translation [1]{[#1]}%
\providecommand \BibitemOpen [0]{}%
\providecommand \bibitemStop [0]{}%
\providecommand \bibitemNoStop [0]{.\EOS\space}%
\providecommand \EOS [0]{\spacefactor3000\relax}%
\providecommand \BibitemShut  [1]{\csname bibitem#1\endcsname}%
\let\auto@bib@innerbib\@empty
%</preamble>
\bibitem [{\citenamefont {Fickett}\ and\ \citenamefont
  {Davis}(1979)}]{Fickett1979}%
  \BibitemOpen
  \bibfield  {author} {\bibinfo {author} {\bibfnamefont {W.}~\bibnamefont
  {Fickett}}\ and\ \bibinfo {author} {\bibfnamefont {W.}~\bibnamefont
  {Davis}},\ }\href@noop {} {\emph {\bibinfo {title} {Detonation: Theory and
  Experiment}}}\ (\bibinfo  {publisher} {Dover Publications},\ \bibinfo {year}
  {1979})\BibitemShut {NoStop}%
\bibitem [{\citenamefont {Lee}(2008)}]{Lee2008}%
  \BibitemOpen
  \bibfield  {author} {\bibinfo {author} {\bibfnamefont {J.}~\bibnamefont
  {Lee}},\ }\href@noop {} {\emph {\bibinfo {title} {The Detonation
  Phenomenon}}}\ (\bibinfo  {publisher} {Cambridge University Press},\ \bibinfo
  {year} {2008})\BibitemShut {NoStop}%
\bibitem [{\citenamefont {Lee}\ and\ \citenamefont
  {Radulescu}(2005)}]{LeeRadulescu2005}%
  \BibitemOpen
  \bibfield  {author} {\bibinfo {author} {\bibfnamefont {J.}~\bibnamefont
  {Lee}}\ and\ \bibinfo {author} {\bibfnamefont {M.}~\bibnamefont
  {Radulescu}},\ }\href@noop {} {\bibfield  {journal} {\bibinfo  {journal}
  {Combustion, Explosion, and Shock Waves}\ }\textbf {\bibinfo {volume} {41}},\
  \bibinfo {pages} {745} (\bibinfo {year} {2005})}\BibitemShut {NoStop}%
\bibitem [{\citenamefont {Radulescu}\ \emph {et~al.}(2007)\citenamefont
  {Radulescu}, \citenamefont {Sharpe}, \citenamefont {Law},\ and\ \citenamefont
  {Lee}}]{Radulescu2007JFM}%
  \BibitemOpen
  \bibfield  {author} {\bibinfo {author} {\bibfnamefont {M.}~\bibnamefont
  {Radulescu}}, \bibinfo {author} {\bibfnamefont {G.}~\bibnamefont {Sharpe}},
  \bibinfo {author} {\bibfnamefont {C.}~\bibnamefont {Law}}, \ and\ \bibinfo
  {author} {\bibfnamefont {J.}~\bibnamefont {Lee}},\ }\href@noop {} {\bibfield
  {journal} {\bibinfo  {journal} {Journal of Fluid Mechanics}\ }\textbf
  {\bibinfo {volume} {580}},\ \bibinfo {pages} {31} (\bibinfo {year}
  {2007})}\BibitemShut {NoStop}%
\bibitem [{\citenamefont {Kiyanda}\ and\ \citenamefont
  {Higgins}(2013)}]{Kiyanda2013}%
  \BibitemOpen
  \bibfield  {author} {\bibinfo {author} {\bibfnamefont {C.}~\bibnamefont
  {Kiyanda}}\ and\ \bibinfo {author} {\bibfnamefont {A.}~\bibnamefont
  {Higgins}},\ }\href {\doibase 10.1007/s00193-012-0413-8} {\bibfield
  {journal} {\bibinfo  {journal} {Shock Waves}\ }\textbf {\bibinfo {volume}
  {23}},\ \bibinfo {pages} {115} (\bibinfo {year} {2013})}\BibitemShut
  {NoStop}%
\bibitem [{\citenamefont {Taki}\ and\ \citenamefont
  {Fujiwara}(1978)}]{Taki1978}%
  \BibitemOpen
  \bibfield  {author} {\bibinfo {author} {\bibfnamefont {S.}~\bibnamefont
  {Taki}}\ and\ \bibinfo {author} {\bibfnamefont {T.}~\bibnamefont
  {Fujiwara}},\ }\href@noop {} {\bibfield  {journal} {\bibinfo  {journal} {AIAA
  Journal}\ }\textbf {\bibinfo {volume} {16}},\ \bibinfo {pages} {73} (\bibinfo
  {year} {1978})}\BibitemShut {NoStop}%
\bibitem [{\citenamefont {Oran}\ \emph {et~al.}(1982)\citenamefont {Oran},
  \citenamefont {Young}, \citenamefont {Boris}, \citenamefont {Picone},\ and\
  \citenamefont {Edwards}}]{Oran1982}%
  \BibitemOpen
  \bibfield  {author} {\bibinfo {author} {\bibfnamefont {E.}~\bibnamefont
  {Oran}}, \bibinfo {author} {\bibfnamefont {T.}~\bibnamefont {Young}},
  \bibinfo {author} {\bibfnamefont {J.}~\bibnamefont {Boris}}, \bibinfo
  {author} {\bibfnamefont {J.}~\bibnamefont {Picone}}, \ and\ \bibinfo {author}
  {\bibfnamefont {D.}~\bibnamefont {Edwards}},\ }\href {\doibase
  http://dx.doi.org/10.1016/S0082-0784(82)80231-2} {\bibfield  {journal}
  {\bibinfo  {journal} {Symposium (International) on Combustion}\ }\textbf
  {\bibinfo {volume} {19}},\ \bibinfo {pages} {573 } (\bibinfo {year}
  {1982})}\BibitemShut {NoStop}%
\bibitem [{\citenamefont {Mahmoudi}\ and\ \citenamefont
  {Mazaheri}(2011)}]{Mahmoudi2011}%
  \BibitemOpen
  \bibfield  {author} {\bibinfo {author} {\bibfnamefont {Y.}~\bibnamefont
  {Mahmoudi}}\ and\ \bibinfo {author} {\bibfnamefont {K.}~\bibnamefont
  {Mazaheri}},\ }\href {\doibase http://dx.doi.org/10.1016/j.proci.2010.07.083}
  {\bibfield  {journal} {\bibinfo  {journal} {Proceedings of the Combustion
  Institute}\ }\textbf {\bibinfo {volume} {33}},\ \bibinfo {pages} {2187 }
  (\bibinfo {year} {2011})}\BibitemShut {NoStop}%
\bibitem [{\citenamefont {Mazaheri}\ \emph {et~al.}(2012)\citenamefont
  {Mazaheri}, \citenamefont {Mahmoudi},\ and\ \citenamefont
  {Radulescu}}]{Mazaheri2012}%
  \BibitemOpen
  \bibfield  {author} {\bibinfo {author} {\bibfnamefont {K.}~\bibnamefont
  {Mazaheri}}, \bibinfo {author} {\bibfnamefont {Y.}~\bibnamefont {Mahmoudi}},
  \ and\ \bibinfo {author} {\bibfnamefont {M.}~\bibnamefont {Radulescu}},\
  }\href {\doibase http://dx.doi.org/10.1016/j.combustflame.2012.01.024}
  {\bibfield  {journal} {\bibinfo  {journal} {Combustion and Flame}\ }\textbf
  {\bibinfo {volume} {159}},\ \bibinfo {pages} {2138 } (\bibinfo {year}
  {2012})}\BibitemShut {NoStop}%
\bibitem [{\citenamefont {Lu}\ and\ \citenamefont {Braun}(2014)}]{Lu2014RDE}%
  \BibitemOpen
  \bibfield  {author} {\bibinfo {author} {\bibfnamefont {F.}~\bibnamefont
  {Lu}}\ and\ \bibinfo {author} {\bibfnamefont {E.}~\bibnamefont {Braun}},\
  }\href@noop {} {\bibfield  {journal} {\bibinfo  {journal} {Journal of
  Propulsion and Power}\ }\textbf {\bibinfo {volume} {30}},\ \bibinfo {pages}
  {1125} (\bibinfo {year} {2014})}\BibitemShut {NoStop}%
\bibitem [{\citenamefont {Higgins}(2012)}]{HigginsChapter2}%
  \BibitemOpen
  \bibfield  {author} {\bibinfo {author} {\bibfnamefont {A.}~\bibnamefont
  {Higgins}},\ }in\ \href@noop {} {\emph {\bibinfo {booktitle} {Shock Waves
  Science and Technology Library}}},\ Vol.~\bibinfo {volume} {6},\ \bibinfo
  {editor} {edited by\ \bibinfo {editor} {\bibfnamefont {F.}~\bibnamefont
  {Zhang}}}\ (\bibinfo  {publisher} {Springer Berlin Heidelberg},\ \bibinfo
  {year} {2012})\ pp.\ \bibinfo {pages} {33--105}\BibitemShut {NoStop}%
\bibitem [{\citenamefont {Mi}\ \emph {et~al.}(2017{\natexlab{a}})\citenamefont
  {Mi}, \citenamefont {Timofeev},\ and\ \citenamefont {Higgins}}]{Mi2016JFM}%
  \BibitemOpen
  \bibfield  {author} {\bibinfo {author} {\bibfnamefont {X.}~\bibnamefont
  {Mi}}, \bibinfo {author} {\bibfnamefont {E.}~\bibnamefont {Timofeev}}, \ and\
  \bibinfo {author} {\bibfnamefont {A.}~\bibnamefont {Higgins}},\ }\href@noop
  {} {\bibfield  {journal} {\bibinfo  {journal} {Journal of Fluid Mechanics}\ }
  (\bibinfo {year} {2017}{\natexlab{a}})},\ \bibinfo {note} {accepted,
  https://arxiv.org/abs/1608.07665}\BibitemShut {NoStop}%
\bibitem [{\citenamefont {Lee}\ and\ \citenamefont
  {Stewart}(1990)}]{LeeStewart1990}%
  \BibitemOpen
  \bibfield  {author} {\bibinfo {author} {\bibfnamefont {H.}~\bibnamefont
  {Lee}}\ and\ \bibinfo {author} {\bibfnamefont {D.}~\bibnamefont {Stewart}},\
  }\href@noop {} {\bibfield  {journal} {\bibinfo  {journal} {Journal of Fluid
  Mechanics}\ }\textbf {\bibinfo {volume} {216}},\ \bibinfo {pages} {103}
  (\bibinfo {year} {1990})}\BibitemShut {NoStop}%
\bibitem [{\citenamefont {Short}\ and\ \citenamefont
  {Stewart}(1998)}]{Short1998}%
  \BibitemOpen
  \bibfield  {author} {\bibinfo {author} {\bibfnamefont {M.}~\bibnamefont
  {Short}}\ and\ \bibinfo {author} {\bibfnamefont {D.}~\bibnamefont
  {Stewart}},\ }\href@noop {} {\bibfield  {journal} {\bibinfo  {journal}
  {Journal of Fluid Mechanics}\ }\textbf {\bibinfo {volume} {368}},\ \bibinfo
  {pages} {229} (\bibinfo {year} {1998})}\BibitemShut {NoStop}%
\bibitem [{\citenamefont {Toro}(2009)}]{Toro2009}%
  \BibitemOpen
  \bibfield  {author} {\bibinfo {author} {\bibfnamefont {E.~F.}\ \bibnamefont
  {Toro}},\ }\href@noop {} {\emph {\bibinfo {title} {Riemann Solvers and
  Numerical Methods for Fluid Dynamics}}},\ \bibinfo {edition} {3rd}\ ed.\
  (\bibinfo  {publisher} {Springer},\ \bibinfo {year} {2009})\BibitemShut
  {NoStop}%
\bibitem [{\citenamefont {Strang}(1968)}]{Strang1968}%
  \BibitemOpen
  \bibfield  {author} {\bibinfo {author} {\bibfnamefont {G.}~\bibnamefont
  {Strang}},\ }\href@noop {} {\bibfield  {journal} {\bibinfo  {journal} {SIAM
  Journal on Numerical Analysis}\ }\textbf {\bibinfo {volume} {5}},\ \bibinfo
  {pages} {506} (\bibinfo {year} {1968})}\BibitemShut {NoStop}%
\bibitem [{\citenamefont {Morgan}(2013)}]{Morgan2013}%
  \BibitemOpen
  \bibfield  {author} {\bibinfo {author} {\bibfnamefont {G.}~\bibnamefont
  {Morgan}},\ }\href@noop {} {\emph {\bibinfo {title} {The Euler Equations with
  a Single-Step Arrhenius Reaction}}},\ \bibinfo {type} {Tech. Rep.}\ (\bibinfo
   {institution} {University of Cambridge},\ \bibinfo {year}
  {2013})\BibitemShut {NoStop}%
\bibitem [{\citenamefont {Kiyanda}\ \emph {et~al.}(2015)\citenamefont
  {Kiyanda}, \citenamefont {Morgan}, \citenamefont {Nikiforakis},\ and\
  \citenamefont {Ng}}]{KiyandaNg2015}%
  \BibitemOpen
  \bibfield  {author} {\bibinfo {author} {\bibfnamefont {C.}~\bibnamefont
  {Kiyanda}}, \bibinfo {author} {\bibfnamefont {G.}~\bibnamefont {Morgan}},
  \bibinfo {author} {\bibfnamefont {N.}~\bibnamefont {Nikiforakis}}, \ and\
  \bibinfo {author} {\bibfnamefont {H.}~\bibnamefont {Ng}},\ }\href@noop {}
  {\bibfield  {journal} {\bibinfo  {journal} {Journal of Visualization}\
  }\textbf {\bibinfo {volume} {18}},\ \bibinfo {pages} {273} (\bibinfo {year}
  {2015})}\BibitemShut {NoStop}%
\bibitem [{\citenamefont {Sow}\ \emph {et~al.}(2014)\citenamefont {Sow},
  \citenamefont {Chinnayya},\ and\ \citenamefont {Hadjadj}}]{Sow2014JFM}%
  \BibitemOpen
  \bibfield  {author} {\bibinfo {author} {\bibfnamefont {A.}~\bibnamefont
  {Sow}}, \bibinfo {author} {\bibfnamefont {A.}~\bibnamefont {Chinnayya}}, \
  and\ \bibinfo {author} {\bibfnamefont {A.}~\bibnamefont {Hadjadj}},\
  }\href@noop {} {\bibfield  {journal} {\bibinfo  {journal} {Journal of Fluid
  Mechanics}\ }\textbf {\bibinfo {volume} {743}},\ \bibinfo {pages} {503}
  (\bibinfo {year} {2014})}\BibitemShut {NoStop}%
\bibitem [{\citenamefont {Goroshin}\ \emph {et~al.}(1998)\citenamefont
  {Goroshin}, \citenamefont {Lee},\ and\ \citenamefont
  {Shoshin}}]{GoroshinLeeShoshin1998}%
  \BibitemOpen
  \bibfield  {author} {\bibinfo {author} {\bibfnamefont {S.}~\bibnamefont
  {Goroshin}}, \bibinfo {author} {\bibfnamefont {J.}~\bibnamefont {Lee}}, \
  and\ \bibinfo {author} {\bibfnamefont {Y.}~\bibnamefont {Shoshin}},\
  }\href@noop {} {\bibfield  {journal} {\bibinfo  {journal} {Proceedings of the
  Combustion Institute}\ }\textbf {\bibinfo {volume} {27}},\ \bibinfo {pages}
  {743} (\bibinfo {year} {1998})}\BibitemShut {NoStop}%
\bibitem [{\citenamefont {Beck}\ and\ \citenamefont
  {Volpert}(2003)}]{Beck2003}%
  \BibitemOpen
  \bibfield  {author} {\bibinfo {author} {\bibfnamefont {J.}~\bibnamefont
  {Beck}}\ and\ \bibinfo {author} {\bibfnamefont {V.}~\bibnamefont {Volpert}},\
  }\href@noop {} {\bibfield  {journal} {\bibinfo  {journal} {Physica D:
  Nonlinear Phenomena}\ }\textbf {\bibinfo {volume} {182}},\ \bibinfo {pages}
  {86} (\bibinfo {year} {2003})}\BibitemShut {NoStop}%
\bibitem [{\citenamefont {Tang}\ \emph {et~al.}(2009)\citenamefont {Tang},
  \citenamefont {Higgins},\ and\ \citenamefont {Goroshin}}]{Tang2009CTM}%
  \BibitemOpen
  \bibfield  {author} {\bibinfo {author} {\bibfnamefont {F.}~\bibnamefont
  {Tang}}, \bibinfo {author} {\bibfnamefont {A.}~\bibnamefont {Higgins}}, \
  and\ \bibinfo {author} {\bibfnamefont {S.}~\bibnamefont {Goroshin}},\
  }\href@noop {} {\bibfield  {journal} {\bibinfo  {journal} {Combustion Theory
  and Modelling}\ }\textbf {\bibinfo {volume} {13}},\ \bibinfo {pages} {319}
  (\bibinfo {year} {2009})}\BibitemShut {NoStop}%
\bibitem [{\citenamefont {Goroshin}\ \emph {et~al.}(2011)\citenamefont
  {Goroshin}, \citenamefont {Tang},\ and\ \citenamefont
  {Higgins}}]{Goroshin2011PRE}%
  \BibitemOpen
  \bibfield  {author} {\bibinfo {author} {\bibfnamefont {S.}~\bibnamefont
  {Goroshin}}, \bibinfo {author} {\bibfnamefont {F.}~\bibnamefont {Tang}}, \
  and\ \bibinfo {author} {\bibfnamefont {A.}~\bibnamefont {Higgins}},\
  }\href@noop {} {\bibfield  {journal} {\bibinfo  {journal} {Physical Review
  E}\ }\textbf {\bibinfo {volume} {84}},\ \bibinfo {pages} {027301} (\bibinfo
  {year} {2011})}\BibitemShut {NoStop}%
\bibitem [{\citenamefont {Tang}\ \emph {et~al.}(2012)\citenamefont {Tang},
  \citenamefont {Higgins},\ and\ \citenamefont {Goroshin}}]{Tang2011PRE}%
  \BibitemOpen
  \bibfield  {author} {\bibinfo {author} {\bibfnamefont {F.}~\bibnamefont
  {Tang}}, \bibinfo {author} {\bibfnamefont {A.}~\bibnamefont {Higgins}}, \
  and\ \bibinfo {author} {\bibfnamefont {S.}~\bibnamefont {Goroshin}},\
  }\href@noop {} {\bibfield  {journal} {\bibinfo  {journal} {Physical Review
  E}\ }\textbf {\bibinfo {volume} {85}},\ \bibinfo {pages} {036311} (\bibinfo
  {year} {2012})}\BibitemShut {NoStop}%
\bibitem [{\citenamefont {Mi}\ \emph {et~al.}(2017{\natexlab{b}})\citenamefont
  {Mi}, \citenamefont {Higgins}, \citenamefont {Goroshin},\ and\ \citenamefont
  {Bergthorson}}]{Mi2016PROCI}%
  \BibitemOpen
  \bibfield  {author} {\bibinfo {author} {\bibfnamefont {X.}~\bibnamefont
  {Mi}}, \bibinfo {author} {\bibfnamefont {A.}~\bibnamefont {Higgins}},
  \bibinfo {author} {\bibfnamefont {S.}~\bibnamefont {Goroshin}}, \ and\
  \bibinfo {author} {\bibfnamefont {J.}~\bibnamefont {Bergthorson}},\
  }\href@noop {} {\bibfield  {journal} {\bibinfo  {journal} {Proceedings of the
  Combustion Institute}\ }\textbf {\bibinfo {volume} {36}},\ \bibinfo {pages}
  {2359} (\bibinfo {year} {2017}{\natexlab{b}})}\BibitemShut {NoStop}%
\bibitem [{\citenamefont {Wright}\ \emph {et~al.}(2016)\citenamefont {Wright},
  \citenamefont {Higgins},\ and\ \citenamefont {Goroshin}}]{Wright2016}%
  \BibitemOpen
  \bibfield  {author} {\bibinfo {author} {\bibfnamefont {A.}~\bibnamefont
  {Wright}}, \bibinfo {author} {\bibfnamefont {A.}~\bibnamefont {Higgins}}, \
  and\ \bibinfo {author} {\bibfnamefont {S.}~\bibnamefont {Goroshin}},\
  }\href@noop {} {\bibfield  {journal} {\bibinfo  {journal} {Combustion Science
  and Technology}\ }\textbf {\bibinfo {volume} {188}},\ \bibinfo {pages} {2178}
  (\bibinfo {year} {2016})}\BibitemShut {NoStop}%
\bibitem [{\citenamefont {Hishida}\ \emph {et~al.}(2009)\citenamefont
  {Hishida}, \citenamefont {Fujiwara},\ and\ \citenamefont
  {Wolanski}}]{Hishida2009}%
  \BibitemOpen
  \bibfield  {author} {\bibinfo {author} {\bibfnamefont {M.}~\bibnamefont
  {Hishida}}, \bibinfo {author} {\bibfnamefont {T.}~\bibnamefont {Fujiwara}}, \
  and\ \bibinfo {author} {\bibfnamefont {P.}~\bibnamefont {Wolanski}},\ }\href
  {\doibase 10.1007/s00193-008-0178-2} {\bibfield  {journal} {\bibinfo
  {journal} {Shock Waves}\ }\textbf {\bibinfo {volume} {19}},\ \bibinfo {pages}
  {1} (\bibinfo {year} {2009})}\BibitemShut {NoStop}%
\bibitem [{\citenamefont {Fujii}\ \emph {et~al.}(2017)\citenamefont {Fujii},
  \citenamefont {Kumazawa}, \citenamefont {Matsuo}, \citenamefont {Nakagami},
  \citenamefont {Matsuoka},\ and\ \citenamefont {Kasahara}}]{Fujii2016}%
  \BibitemOpen
  \bibfield  {author} {\bibinfo {author} {\bibfnamefont {J.}~\bibnamefont
  {Fujii}}, \bibinfo {author} {\bibfnamefont {Y.}~\bibnamefont {Kumazawa}},
  \bibinfo {author} {\bibfnamefont {A.}~\bibnamefont {Matsuo}}, \bibinfo
  {author} {\bibfnamefont {S.}~\bibnamefont {Nakagami}}, \bibinfo {author}
  {\bibfnamefont {K.}~\bibnamefont {Matsuoka}}, \ and\ \bibinfo {author}
  {\bibfnamefont {J.}~\bibnamefont {Kasahara}},\ }\href@noop {} {\bibfield
  {journal} {\bibinfo  {journal} {Proceedings of the Combustion Institute}\
  }\textbf {\bibinfo {volume} {36}},\ \bibinfo {pages} {2665} (\bibinfo {year}
  {2017})}\BibitemShut {NoStop}%
\bibitem [{\citenamefont {Yao}\ \emph {et~al.}(2016)\citenamefont {Yao},
  \citenamefont {Han}, \citenamefont {Liu},\ and\ \citenamefont
  {Wang}}]{Yao2016}%
  \BibitemOpen
  \bibfield  {author} {\bibinfo {author} {\bibfnamefont {S.}~\bibnamefont
  {Yao}}, \bibinfo {author} {\bibfnamefont {X.}~\bibnamefont {Han}}, \bibinfo
  {author} {\bibfnamefont {Y.}~\bibnamefont {Liu}}, \ and\ \bibinfo {author}
  {\bibfnamefont {J.}~\bibnamefont {Wang}},\ }\href {\doibase
  http://dx.doi.org/10.1007/s00193-016-0692-6} {\bibfield  {journal} {\bibinfo
  {journal} {Shock Waves}\ } (\bibinfo {year} {2016}),\
  http://dx.doi.org/10.1007/s00193-016-0692-6},\ \bibinfo {note}
  {online}\BibitemShut {NoStop}%
\bibitem [{\citenamefont {Kindracki}\ \emph {et~al.}(2011)\citenamefont
  {Kindracki}, \citenamefont {Wola{\'{n}}ski},\ and\ \citenamefont
  {Gut}}]{Kindracki2011}%
  \BibitemOpen
  \bibfield  {author} {\bibinfo {author} {\bibfnamefont {J.}~\bibnamefont
  {Kindracki}}, \bibinfo {author} {\bibfnamefont {P.}~\bibnamefont
  {Wola{\'{n}}ski}}, \ and\ \bibinfo {author} {\bibfnamefont {Z.}~\bibnamefont
  {Gut}},\ }\href {\doibase 10.1007/s00193-011-0298-y} {\bibfield  {journal}
  {\bibinfo  {journal} {Shock Waves}\ }\textbf {\bibinfo {volume} {21}},\
  \bibinfo {pages} {75} (\bibinfo {year} {2011})}\BibitemShut {NoStop}%
\bibitem [{\citenamefont {Lin}\ \emph {et~al.}(2015)\citenamefont {Lin},
  \citenamefont {Zhou}, \citenamefont {Liu},\ and\ \citenamefont
  {Lin}}]{Lin2015}%
  \BibitemOpen
  \bibfield  {author} {\bibinfo {author} {\bibfnamefont {W.}~\bibnamefont
  {Lin}}, \bibinfo {author} {\bibfnamefont {J.}~\bibnamefont {Zhou}}, \bibinfo
  {author} {\bibfnamefont {S.}~\bibnamefont {Liu}}, \ and\ \bibinfo {author}
  {\bibfnamefont {Z.}~\bibnamefont {Lin}},\ }\href {\doibase
  http://dx.doi.org/10.1016/j.expthermflusci.2014.11.017} {\bibfield  {journal}
  {\bibinfo  {journal} {Experimental Thermal and Fluid Science}\ }\textbf
  {\bibinfo {volume} {62}},\ \bibinfo {pages} {122 } (\bibinfo {year}
  {2015})}\BibitemShut {NoStop}%
\bibitem [{\citenamefont {Rankin}\ \emph {et~al.}(2016)\citenamefont {Rankin},
  \citenamefont {Fugger}, \citenamefont {Richardson}, \citenamefont {Cho},
  \citenamefont {Hoke}, \citenamefont {Caswell}, \citenamefont {Gord},\ and\
  \citenamefont {Schauer}}]{Rankin2016}%
  \BibitemOpen
  \bibfield  {author} {\bibinfo {author} {\bibfnamefont {B.}~\bibnamefont
  {Rankin}}, \bibinfo {author} {\bibfnamefont {C.}~\bibnamefont {Fugger}},
  \bibinfo {author} {\bibfnamefont {D.}~\bibnamefont {Richardson}}, \bibinfo
  {author} {\bibfnamefont {K.}~\bibnamefont {Cho}}, \bibinfo {author}
  {\bibfnamefont {J.}~\bibnamefont {Hoke}}, \bibinfo {author} {\bibfnamefont
  {A.}~\bibnamefont {Caswell}}, \bibinfo {author} {\bibfnamefont
  {J.}~\bibnamefont {Gord}}, \ and\ \bibinfo {author} {\bibfnamefont
  {F.}~\bibnamefont {Schauer}},\ }in\ \href@noop {} {\emph {\bibinfo
  {booktitle} {54th AIAA Aerospace Sciences Meeting}}},\ \bibinfo {series and
  number} {\bibinfo {number} {2016-1198}}\ (\bibinfo {year} {2016})\BibitemShut
  {NoStop}%
\bibitem [{\citenamefont {Vasil'ev}\ and\ \citenamefont
  {Nikolaev}(1978)}]{Vasiliev1978}%
  \BibitemOpen
  \bibfield  {author} {\bibinfo {author} {\bibfnamefont {A.}~\bibnamefont
  {Vasil'ev}}\ and\ \bibinfo {author} {\bibfnamefont {Y.}~\bibnamefont
  {Nikolaev}},\ }\href {\doibase
  http://dx.doi.org/10.1016/0094-5765(78)90004-8} {\bibfield  {journal}
  {\bibinfo  {journal} {Acta Astronautica}\ }\textbf {\bibinfo {volume} {5}},\
  \bibinfo {pages} {983} (\bibinfo {year} {1978})}\BibitemShut {NoStop}%
\bibitem [{\citenamefont {Short}\ and\ \citenamefont
  {Quirk}(1997)}]{ShortQuirk1997}%
  \BibitemOpen
  \bibfield  {author} {\bibinfo {author} {\bibfnamefont {M.}~\bibnamefont
  {Short}}\ and\ \bibinfo {author} {\bibfnamefont {J.}~\bibnamefont {Quirk}},\
  }\href@noop {} {\bibfield  {journal} {\bibinfo  {journal} {Journal of Fluid
  Mechanics}\ }\textbf {\bibinfo {volume} {339}},\ \bibinfo {pages} {89}
  (\bibinfo {year} {1997})}\BibitemShut {NoStop}%
\bibitem [{\citenamefont {Ng}\ and\ \citenamefont {Lee}(2003)}]{NgLee2003}%
  \BibitemOpen
  \bibfield  {author} {\bibinfo {author} {\bibfnamefont {H.}~\bibnamefont
  {Ng}}\ and\ \bibinfo {author} {\bibfnamefont {J.~H.}\ \bibnamefont {Lee}},\
  }\href@noop {} {\bibfield  {journal} {\bibinfo  {journal} {Journal of Fluid
  Mechanics}\ }\textbf {\bibinfo {volume} {476}},\ \bibinfo {pages} {179}
  (\bibinfo {year} {2003})}\BibitemShut {NoStop}%
\end{thebibliography}%

\end{document}